\pdfoutput=1
\documentclass[iop]{emulateapj}
\usepackage{amsmath}

\slugcomment{}

\shorttitle{The embedded ring-like feature and star formation activities in G35.673-00.847}
\shortauthors{L.~K. Dewangan et al.}
\begin{document}

\title{The embedded ring-like feature and star formation activities in G35.673-00.847}
\author{L.~K. Dewangan\altaffilmark{1}, R. Devaraj\altaffilmark{2}, and D.~K. Ojha\altaffilmark{3}}
\email{lokeshd@prl.res.in}
\altaffiltext{1}{Physical Research Laboratory, Navrangpura, Ahmedabad - 380 009, India.}
\altaffiltext{2}{Instituto Nacional de Astrof\'{\i}sica, \'{O}ptica y Electr\'{o}nica, Luis Enrique Erro \# 1, Tonantzintla, Puebla, M\'{e}xico C.P. 72840.}
\altaffiltext{3}{Department of Astronomy and Astrophysics, Tata Institute of Fundamental Research, Homi Bhabha Road, Mumbai 400 005, India.}
\begin{abstract}
We present a multi-wavelength study to probe the star formation (SF) process 
in the molecular cloud linked with the G35.673-00.847 site (hereafter MCG35.6), which is traced in a velocity range of 53--62 km s$^{-1}$. 
Multi-wavelength images reveal a semi-ring-like feature (associated with ionized gas emission) and an embedded face-on ring-like feature (without the NVSS 1.4 GHz radio emission; where 1$\sigma$ $\sim$0.45 mJy beam$^{-1}$) in the MCG35.6. 
The semi-ring-like feature is originated by the ionizing feedback from a star with spectral type B0.5V--B0V. 
The central region of the ring-like feature does not contain detectable ionized gas emission, indicating that the ring-like feature is unlikely to be produced by the ionizing feedback from a massive star. Several embedded {\it Herschel} clumps and young stellar objects (YSOs) are identified in the MCG35.6, tracing the ongoing SF activities within the cloud. 
The polarization information from the {\it Planck} and GPIPS data trace the plane-of-sky magnetic field, which is oriented 
parallel to the major axis of the ring-like feature. 
At least five clumps (having $M_\mathrm{clump}$ $\sim$740--1420 M$_{\odot}$) seem to be distributed in 
an almost regularly spaced manner along the ring-like feature and contain noticeable YSOs.
Based on the analysis of the polarization and molecular line data, three subregions containing 
the clumps are found to be magnetically supercritical in the ring-like feature.
Altogether, the existence of the ring-like feature and the SF activities on 
its edges can be explained by the magnetic field mediated process as 
simulated by Li \& Nakamura (2002).
 \end{abstract}
\keywords{dust, extinction -- HII regions -- ISM: clouds -- ISM: individual objects (G35.673-00.847) -- ISM: Magnetic fields -- stars: formation} 
\section{Introduction}
\label{sec:intro}
In recent years, space-based infrared observations have demonstrated that the infrared structures (such as, bubbles, ring-like features, and shells) 
are often associated with young stellar clusters, and are commonly observed in Galactic molecular clouds \citep[e.g.][]{churchwell06,churchwell07,anderson09,beaumont10,deharveng10}. 
\citet{churchwell06} reported that 25\% of 322 mid-infrared (MIR) bubbles, distributed outside 10$\degr$ of the Galactic center, are associated with known H\,{\sc ii} regions detected at radio wavelengths. 
Furthermore, using the same catalog given in \citet{churchwell06}, \citet{deharveng10} extended this 
information to 86\% out of 102 bubbles through the detection of the 20 cm radio continuum emission. 
Hence, based on the detection/non-detection of the centimeter radio continuum emission at the sensitivity of the observed radio map, the infrared structures are seen either associated with H\,{\sc ii} regions or 
devoid of H\,{\sc ii} regions. 
The infrared structures associated with H\,{\sc ii} regions are promising sites to investigate the feedback processes of massive OB stars \citep[e.g.][]{deharveng10,kendrew12,thompson12,dewangan16,dewangan17a}. 
The photoionized gas and/or stellar winds associated with the OB stars are often 
considered as the major contributor for the origin of the infrared structures. 
Another possible mechanism responsible for the infrared structures could be supernovae explosions. 
On the other hand, the existence of a ring-like morphology without H\,{\sc ii} regions is unlikely to be 
explained by the feedback of OB stars.
The magnetic field mediated scenario has predicted the formation of ring-like features in magnetically subcritical clouds (e.g. \citet{nakamura02}; \citet{li02} and references therein). 
These authors also examined the evolution of these subcritical clouds/ring-like features, 
and found that a subcritical cloud fragments into multiple magnetically supercritical cores/clumps, 
where the birth of young stellar populations is expected. 
Recently, \citet{pavel12} examined the interaction of H\,{\sc ii} region driven Galactic bubbles with the Galactic magnetic field, and found that external magnetic fields are 
important during the earliest phases of the evolution of H\,{\sc ii} region driven Galactic bubbles. However, to our knowledge, a detailed observational study of the influence of the magnetic field 
on bubble/ring-like structures is still limited in the literature \citep[e.g.,][]{chen17}. 
Interestingly, the presence of infrared structures with H\,{\sc ii} regions and without 
H\,{\sc ii} regions in a single star-forming site can offer the possibility to explore the numerous complex physical processes involved in star formation. 
A detailed multi-wavelength study of such site can provide important observational evidences which can directly lead to examine the 
existing theoretical scenarios concerning the formation of stellar clusters and ring-like features not associated with H\,{\sc ii} regions.

The IRAS 18569+0159/G35.673-00.847 region (hereafter G35.6) is located at a distance of 3.7 kpc \citep{paron11} and is a poorly explored star-forming site. Using the radio continuum 1.4 GHz map from the NRAO VLA Sky Survey \citep[NVSS; 1.4 GHz;][]{condon98}, 
\citet{paron11} studied the H\,{\sc ii} region associated with the G35.6 site (hereafter G35.6 H\,{\sc ii} region; see Figure~\ref{fg2}a in this paper). Based on the MIR and radio continuum emission, the G35.6 site has an almost semi-ring-like appearance with a radius 
of about 1$\farcm$5 ($\sim$1.6 pc at a distance of 3.7 kpc) \citep{paron11}, which is associated 
with the main radio continuum source detected in the region (see Figures~\ref{fg2}a and~\ref{fg2}b). 
In the direction of the G35.6 H\,{\sc ii} region, the velocity of the ionized gas is estimated to be 60.7$\pm$2.6 km s$^{-1}$ \citep{lockman89}.
Using the Galactic Ring Survey \citep[GRS;][]{jackson06} $^{13}$CO (J=1$-$0) line data, 
\citet{anderson09} computed the radial velocity of the molecular gas to be $\sim$57.1 km s$^{-1}$ and designated the molecular cloud associated 
with the G35.6 site (hereafter MCG35.6) as D35.67-0.85. 
Later, using the GRS $^{13}$CO line data, \citet{paron11} reported the presence of a molecular shell (with a mass $M_\mathrm{cloud}$ $\sim$10$^{4}$ M$_{\odot}$) containing several young stellar objects (YSOs, see Figure~\ref{fg2}b). 
The YSOs were identified using the Two Micron All Sky Survey \citep[2MASS; $\lambda$ = 1.25, 1.65, 2.2 $\mu$m;][]{skrutskie06} 
and the Galactic Legacy Infrared Mid-Plane Survey Extraordinaire \citep[GLIMPSE; $\lambda$ = 3.6, 4.5, 5.8, 8.0 $\mu$m;][]{benjamin03} 
photometric data sets. the majority of the $^{13}$CO gas in the molecular shell is found away from the radio continuum emission \citep[see Figure~4 in][]{paron11}. 
In this paper, Figures~\ref{fg2}a and~\ref{fg2}b show the observed ionized and molecular 
features previously reported by \citet{paron11}, and also reveal that the G35.6 H\,{\sc ii} region is 
spatially seen in the Galactic northern side of the molecular shell. 
\citet{paron11} suggested that the G35.6 H\,{\sc ii} region is interacting with its surrounding molecular cloud.  
However, the origin of the infrared and molecular structures seen in the G35.6 site still remains unexplored. The physical conditions in the G35.6 site are yet to be probed. 
Furthermore, the magnetic field properties in the molecular structure have not yet been studied. The investigation of dust continuum clumps and embedded young stellar clusters in the MCG35.6 is also yet to be carried out. 

In this paper, to understand the ongoing physical processes and star formation activity in the G35.6 site, we present a detailed multi-wavelength study of observations 
from near-infrared (NIR) to radio wavelengths. 
The multi-wavelength data sets (including imaging, spectroscopic, and polarimetric observations) are obtained from various Galactic plane surveys. 
A careful investigation of these data sets allows us to probe the distribution of column density, extinction, dust temperature, ionized gas emission, 
kinematics of $^{13}$CO gas, plane-of-sky (POS) magnetic field, and YSOs in the MCG35.6.  

This paper is structured as follows. In Section~\ref{sec:obser}, we discuss the data selection. 
The results of our extensive multi-wavelength data analysis are presented in Section~\ref{sec:data}. 
The implications of our findings concerning the star formation are discussed in Section~\ref{sec:disc}.
Finally, our results are summarized and concluded in Section~\ref{sec:conc}.
\section{Data sets and analysis}
\label{sec:obser}
In this section, we provide a brief description of the adopted multi-wavelength data sets from various 
Galactic plane surveys (see also Table~\ref{ftab1}). In the present work, we selected a field of $\sim$27$'$ $\times$ 20$\farcm5$ ($\sim$29 pc $\times$ 22 pc); centered at $l$ = 35$\degr$.734; $b$ = $-$0$\degr$.914.
\subsection{NIR Data}
We analyzed the photometric NIR data extracted from the United Kingdom Infra-Red Telescope (UKIRT) 
Infrared Deep Sky Survey (UKIDSS) Galactic Plane Survey \citep[GPS; $\lambda$ = 1.25, 1.65, 2.2 $\mu$m;][]{lawrence07} 
sixth archival data release (UKIDSSDR6plus) and the 2MASS catalogs. The UKIDSS observations 
(resolution $\sim$$0\farcs8$) were performed using the Wide Field Camera (WFCAM) at UKIRT. 
The calibration of UKIDSS fluxes was carried out using the 2MASS photometric data. In this work, we extract only reliable NIR photometric data from the catalog. To obtain reliable point sources from the GPS catalog, the conditions suggested by \citet{lucas08} are considered, 
which include the removal of saturated sources, non-stellar sources, and unreliable sources near the sensitivity limits. 
A SQL script \citep[based on][and presented in \citet{dewangan15}]{lucas08} is especially written to implement these criteria.
More information about the selection procedures of the GPS photometry can be found in \citet{dewangan15}.
We notice the UKIDSS saturation near J = 10.3, H = 11.1, and K = 9.6 mag.
Hence, in our final catalog, the magnitudes of the saturated bright UKIDSS sources were obtained from the 2MASS catalog. 
To select good 2MASS photometric quality data, we downloaded only those sources with a photometric 
magnitude error of 0.1 or less in each band.
\subsection{H-band Polarimetry}
To obtain background starlight polarization, NIR H-band (1.6 $\mu$m) linear polarimetric data (resolution $\sim$$1\farcs5$) are extracted from the 
Galactic Plane Infrared Polarization Survey \citep[GPIPS; $\lambda$ = 1.6 $\mu$m;][]{clemens12} Data Release 2 (i.e. DR2). These data sets were observed with the {\it Mimir} instrument, mounted on the 1.8 m Perkins telescope, 
in linear imaging polarimetry mode \citep[see][for more details]{clemens12}. 
To retrieve the most reliable polarimetric data, we selected the sources having 
$P/\sigma_{P}\geq2$, $\sigma_{P}\le5$, and $H \leq 13$ mag (where $P$ is the polarization percentage and $\sigma_p$ is the polarimetric
uncertainty). 
\subsection{{\it Spitzer} and {\it WISE} Data}
We retrieve the photometric images and magnitudes of point sources at 3.6--8.0 $\mu$m 
from the {\it Spitzer}-GLIMPSE survey (resolution $\sim$2$\arcsec$). 
The photometric data are extracted from the GLIMPSE-I Spring '07 highly reliable photometric catalog. 
We also use the Multiband Imaging Photometer for {\it Spitzer} (MIPS) 
Inner Galactic Plane Survey \citep[MIPSGAL; $\lambda$ = 24 $\mu$m;][]{carey05} 24 $\mu$m image (resolution $\sim$6$\arcsec$) and the photometric magnitudes of point-like sources at 24 $\mu$m \citep[from][]{gutermuth15}. 

The archival Wide Field Infrared Survey Explorer (WISE\footnote[1]{WISE is a joint project of the
University of California and the JPL, Caltech, funded by the NASA}; $\lambda$ = 3.4, 4.6, 12, 22 $\mu$m; \citet{wright10}) image at MIR 12 $\mu$m 
(spatial resolution $\sim$6$\arcsec$) is also retrieved.
\subsection{{\it Herschel} Data}
\label{sec:herd}
To probe the far-infrared (FIR) and sub-millimeter (sub-mm) emission, 
{\it Herschel} continuum maps at 70 $\mu$m, 160 $\mu$m, 250 $\mu$m, 350 $\mu$m, and 500 $\mu$m are 
obtained from the {\it Herschel} Space Observatory \citep{pilbratt10,poglitsch10,griffin10,degraauw10} data archives. 
Level2$_{-}$5 processed {\it Herschel} images were downloaded through the {\it Herschel} Interactive Processing Environment \citep[HIPE,][]{ott10}. These observations are part
of the {\it Herschel} Infrared Galactic Plane Survey \citep[Hi-GAL; $\lambda$ = 70, 160, 250, 350, 500 $\mu$m;][]{molinari10} project.  
The plate scales of 70, 160, 250, 350, and 500 $\mu$m images are 3$''$.2, 3$''$.2, 6$''$, 10$''$, and 14$''$ pixel$^{-1}$, respectively.  
The beam sizes of the {\it Herschel} images are 5$\farcs$8, 12$\arcsec$, 18$\arcsec$, 25$\arcsec$, 
and 37$\arcsec$ for 70, 160, 250, 350, and 500 $\mu$m, respectively \citep{poglitsch10,griffin10}. 
The {\it Herschel} images at 70--160 $\mu$m are calibrated in Jy pixel$^{-1}$, while the units of images at 250--500 $\mu$m are MJy sr$^{-1}$.

Following the methods presented in \citet{mallick15}, the {\it Herschel} temperature and column 
density maps are generated using the {\it Herschel} continuum images. These maps are produced from a 
pixel-by-pixel spectral energy distribution (SED) fit with a modified blackbody to the cold dust emission at {\it Herschel} 160--500 $\mu$m \citep[see also][]{dewangan15}. 

A brief step-by-step explanation of the adopted procedures is as follows. 
Before the SED fit, the 160--350 $\mu$m images are converted into the same flux unit (i.e. Jy pixel$^{-1}$) 
and convolved to the lowest angular resolution of the 500 $\mu$m image ($\sim$37$\arcsec$). 
Furthermore, we regrid these images to the pixel size of the 500 $\mu$m image ($\sim$14$\arcsec$). 
These steps are carried out using the plug-in ``Photometric Convolution" and task ``Convert Image Unit" available in the HIPE software. 
Next, the sky background flux level was computed to be 0.211, 0.566, 1.023, and 0.636 Jy pixel$^{-1}$ for the 500, 350, 250, and 
160 $\mu$m images (size of the selected featureless dark region $\sim$4$\farcm$9 $\times$ 6$\farcm$4; 
centered at:  $l$ = 35$\degr$.208; $b$ = $-$1$\degr$.47), respectively. 

In the final step, a modified blackbody is fitted to the observed fluxes on a pixel-by-pixel 
basis to obtain the temperature and column density maps \citep[see equations 8 and 9 in][]{mallick15}. 
The fitting is performed using the four data points for each pixel, retaining the 
dust temperature (T$_{d}$) and the column density ($N(\mathrm H_2)$) 
as free parameters. 
In the analysis, we use a mean molecular weight per hydrogen molecule ($\mu_{H2}$) of 2.8 
\citep{kauffmann08} and an absorption coefficient ($\kappa_\nu$) of 0.1~$(\nu/1000~{\rm GHz})^{\beta}$ cm$^{2}$ g$^{-1}$, 
including a gas-to-dust ratio ($R_t$ ) of 100, with a dust spectral index ($\beta$) of 2 \citep[see][]{hildebrand83}. 
We further discuss these {\it Herschel} maps in Section~\ref{subsec:temp}.
\subsection{{\it Planck} 353 GHz data}
To obtain dust polarized emission, we use the {\it Planck\footnote[2]{{\it Planck} (http://www.esa.int/Planck) is a project of the 
European Space Agency (ESA) with instruments provided by two scientific consortia funded by ESA member states and led by 
Principal Investigators from France and Italy, telescope reflectors provided through a collaboration between ESA and a 
scientific consortium led and funded by Denmark, and additional contributions from NASA (USA).}} polarization data 
\citep[$\nu$ = 353 GHz;][]{planck14} observed with the High Frequency Instrument 
(HFI) at 353 GHz, where the dust polarized emission is brightest. The {\it Planck} intensity and polarization (Stokes parameters $I$, $Q$, and $U$) sky maps are obtained from the 2015 release of 
{\it Planck} legacy archive \citep{planck16a}. The procedures of map-making, calibration, and correction of systematic effects are 
described in \citet{planck16b}. We smooth the maps from an initial angular resolution of 4$'$.9 to 6$\arcmin$ to increase 
the signal-to-noise ratio and to minimize beam depolarization effects. The contribution from the cosmic microwave background (CMB) emission 
is very minimum and has been ignored. 
The intensity and polarization maps are converted from temperature scale of $K_\mathrm{CMB}$ to MJy sr$^{-1}$ using the unit conversion and color correction factor of 246.543 \citep{planck14}. The intensity map is also subtracted from the Galactic zero level offset of 0.0885 MJy sr$^{-1}$ \citep{planck16b}. 
The linear polarization of dust emission integrated along the line-of-sight is measured from the Stokes parameters $I$, $Q$, and $U$ as:
\begin{equation}
P = \sqrt{{Q}^{2}+U^{2}}
\end{equation}
\begin{equation}
p = 100 * \frac{P}{I}
\end{equation}
\begin{equation}
\psi = \frac{1}{2}\tan^{-1}\left({U},{Q}\right).
\end{equation}
where $P$ is the polarization intensity, $p$ is the percentage of dust polarization fraction, and $\psi$ is the polarization angle. 
The {\it Planck} Stokes parameters are in the HEALPix \citep{gorski05} convention having the angle $\psi$ = 0 toward the 
north Galactic pole and positive toward the west (clockwise). We use the IAU convention \citep{hamaker96} by 
changing the {\it Planck} Stokes $U$ map to negative so that the polarization angle is measured anticlockwise relative to 
the north Galactic pole. Since the noise in the maps introduces a positive bias on the measured polarization, 
we debias the polarization following the method given in \citet{plaszczynski14}, and using the full available noise covariance matrix.
\subsection{$^{13}$CO (J=1$-$0) Line Data}
In our selected target region, we probe the molecular gas content in the MCG35.6 using the GRS $^{13}$CO (J=1$-$0) line data. 
The GRS line data have a velocity resolution of 0.21~km\,s$^{-1}$, an angular resolution 
of 45$\arcsec$ with 22$\arcsec$ sampling, a main beam efficiency ($\eta_{\rm mb}$) of $\sim$0.48, 
a velocity coverage of $-$5 to 135~km~s$^{-1}$, and a typical rms sensitivity (1$\sigma$)
of $\approx0.13$~K \citep{jackson06}.  
\subsection{Radio Centimeter Continuum emission}
To trace the ionized gas emission in the G35.6 site, we use the NVSS 1.4 GHz continuum map. The survey covers the sky north of $\delta_{2000}$ = $-$40$\degr$ at 1.4 GHz with a beam of 45$\arcsec$ 
and a nearly uniform sensitivity of $\sim$0.45 mJy/beam \citep{condon98}. 
The Coordinated Radio and Infrared Survey for High-Mass Star Formation \citep[CORNISH;][]{hoare12}
5~GHz (6 cm) radio continuum data (beam size $\sim$1\farcs5) are also utilized in the present work.
\section{Results}
\label{sec:data}
\subsection{Physical environment of G35.6}
\label{sec:hh}
\begin{figure*}
\epsscale{0.55}
\plotone{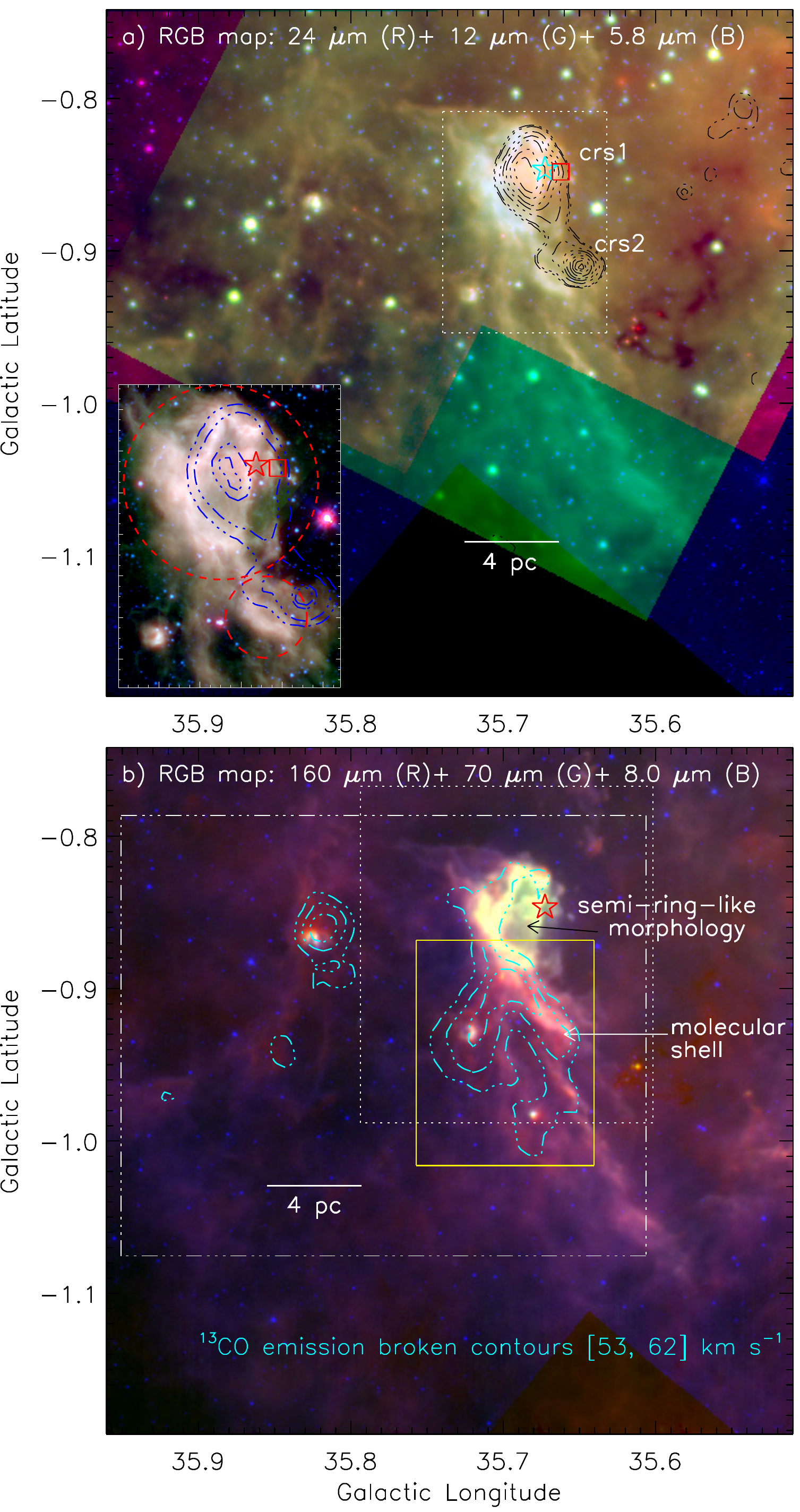}
\caption{\scriptsize A large-scale view of the G35.6 site (size of the selected field $\sim$27$\arcmin$ $\times$ 27$\arcmin$; central
coordinates: $l$ = 35$\degr$.735; $b$ = $-$0$\degr$.967). 
a) The image is the result of the combination of three bands (in log scale): 24.0 $\mu$m in red ({\it Spitzer}), 12.0 $\mu$m in green ({\it WISE}), 
and 5.8 $\mu$m in blue ({\it Spitzer}). 
Contours of NVSS 1.4 GHz radio continuum emission (beam size $\sim$45$\arcsec$) are superimposed with 8, 10, 20, 30, 40, 55, 70, 85, and 95\% of 
the peak value (i.e., 31.7 mJy/beam). Two compact radio sources (i.e., crs1 and crs2) traced in the NVSS 1.4 GHz map are highlighted in the figure. 
The inset on the bottom left shows the H\,{\sc ii} regions in zoomed-in view, 
using a color-composite 
image ({\it WISE} 12.0 $\mu$m (in red), {\it Spitzer} 8.0 $\mu$m (in green), and {\it Spitzer} 5.8 $\mu$m (in blue)) overlaid with the NVSS radio continuum 
emission at 1.4 GHz (see a dashed white box in figure). In the inset, the NVSS contours are 2.5, 6, 20, and 25 mJy/beam. A red square indicates the position of a CORNISH 5~GHz radio source, G035.6624-00.8481 (see also the inset). 
b) The image is the result of the combination of three bands: {\it Herschel} 160 $\mu$m (red), {\it Herschel} 70 $\mu$m (green), 
and {\it Spitzer} 8.0 $\mu$m (blue). The composite map is also overlaid with the $^{13}$CO emission contours integrated over a 
velocity interval from 53 to 62 km s$^{-1}$. The $^{13}$CO contours are shown with the levels of 4.5, 8, and 12 K km s$^{-1}$.
A small dotted box (in white) shows the area investigated by \citet{paron11}.
A semi-ring-like morphology and a molecular shell are also highlighted in the figure \citep[see also][]{paron11}. 
A solid box (in yellow) encompasses the area shown in Figure~\ref{fg3}.
A big dotted-dashed box (in white) encompasses the area shown in Figures~\ref{fg1x} and~\ref{fg2x}a. 
In all the panels, a star symbol indicates the position of IRAS 18569+0159. 
The scale bar corresponding to 4 pc (at a distance of 3.7 kpc) is shown in both the panels.}
\label{fg2}
\end{figure*}

The investigation of the ionized gas, dust (cold and warm), and molecular emission enables us to probe the embedded 
structures present in a given star-forming region. 
In Figure~\ref{fg2}, we show the spatial distribution of dust (warm and cold), molecular gas, and ionized gas toward the G35.6 site. 
The radio continuum data trace the ionized gas emission, while the sub-mm data are sensitive to the cold dust. The MIR and FIR data trace the warm dust.  
Figure~\ref{fg2}a is a three-color composite image made using the MIPSGAL 24 $\mu$m in red, {\it WISE} 12 $\mu$m in green, 
and GLIMPSE 5.8 $\mu$m in blue. 
The color composite image is overlaid with the NVSS radio continuum emission at 1.4 GHz, revealing two 
compact radio sources (designated as crs1 and crs2). 
These radio sources, crs1 and crs2, were previously referred to as NVSS 185929+020334 and 
NVSS 185938+020012, respectively \citep[e.g.,][]{condon98,paron11}.
One of the radio sources, crs1 is seen near the location of IRAS 18569+0159. 
The inset on the bottom left shows a zoomed-in view of the two radio continuum peaks using the {\it WISE} and 
{\it Spitzer} images (12 $\mu$m (red), 8.0 $\mu$m (green), and 5.8 $\mu$m (blue); see Figure~\ref{fg2}a). 
There is an extended emission seen at 5.8--12 $\mu$m.
In the literature, it has been reported that the ionized gas and the MIR/FIR emission (at 12/24/70 $\mu$m) are seen systematically 
correlated in H\,{\sc ii} regions \citep[e.g.][]{deharveng10,paladini12}. In the vicinity of crs1 and crs2, infrared features are also noticeably observed (see broken circles in the inset). 
It is also known that the 5.8--12 $\mu$m bands contain various polycyclic aromatic hydrocarbon (PAH) 
features at 6.2, 7.7, 8.6, and 11.3 $\mu$m (including the continuum). 
The presence of PAH features can be used to trace photodissociation regions (or photon-dominated regions, or PDRs) and enables us to 
infer the impact and influence of the H\,{\sc ii} regions on their surroundings (see broken circles in the inset).
The detection of the CORNISH 5~GHz continuum emission is also indicated by a square in 
Figure~\ref{fg2}a (see also the inset). The quantitative calculations using these radio data 
are performed in Section~\ref{subsec:radio}. 

Figure~\ref{fg2}b shows a three-color composite image made using the {\it Herschel} 160 $\mu$m (in red), {\it Herschel} 70 $\mu$m (in green), 
and GLIMPSE 8.0 $\mu$m (in blue) images. The composite image is overlaid with the $^{13}$CO molecular gas emission. In the direction of our selected region, the $^{13}$CO emission traces the molecular gas in the velocity range of 53--62 km s$^{-1}$. 
The distribution of $^{13}$CO also depicts a molecular shell. 
Furthermore, in the composite map, an almost semi-ring-like feature is prominently observed.
As mentioned earlier, these ionized and molecular features are already known in the literature \citep[e.g.][]{paron11}. 
In the composite map, a small dotted box (in white) highlights the area investigated by \citet{paron11}. 
Additionally, a molecular condensation is also observed in the 
Galactic eastern side of the molecular shell (see Figure~\ref{fg2}b). 
A detailed analysis of the molecular and ionized features is presented 
in Sections~\ref{sec:coem} and~\ref{subsec:radio}, respectively.

\begin{figure*}
\epsscale{1}
\plotone{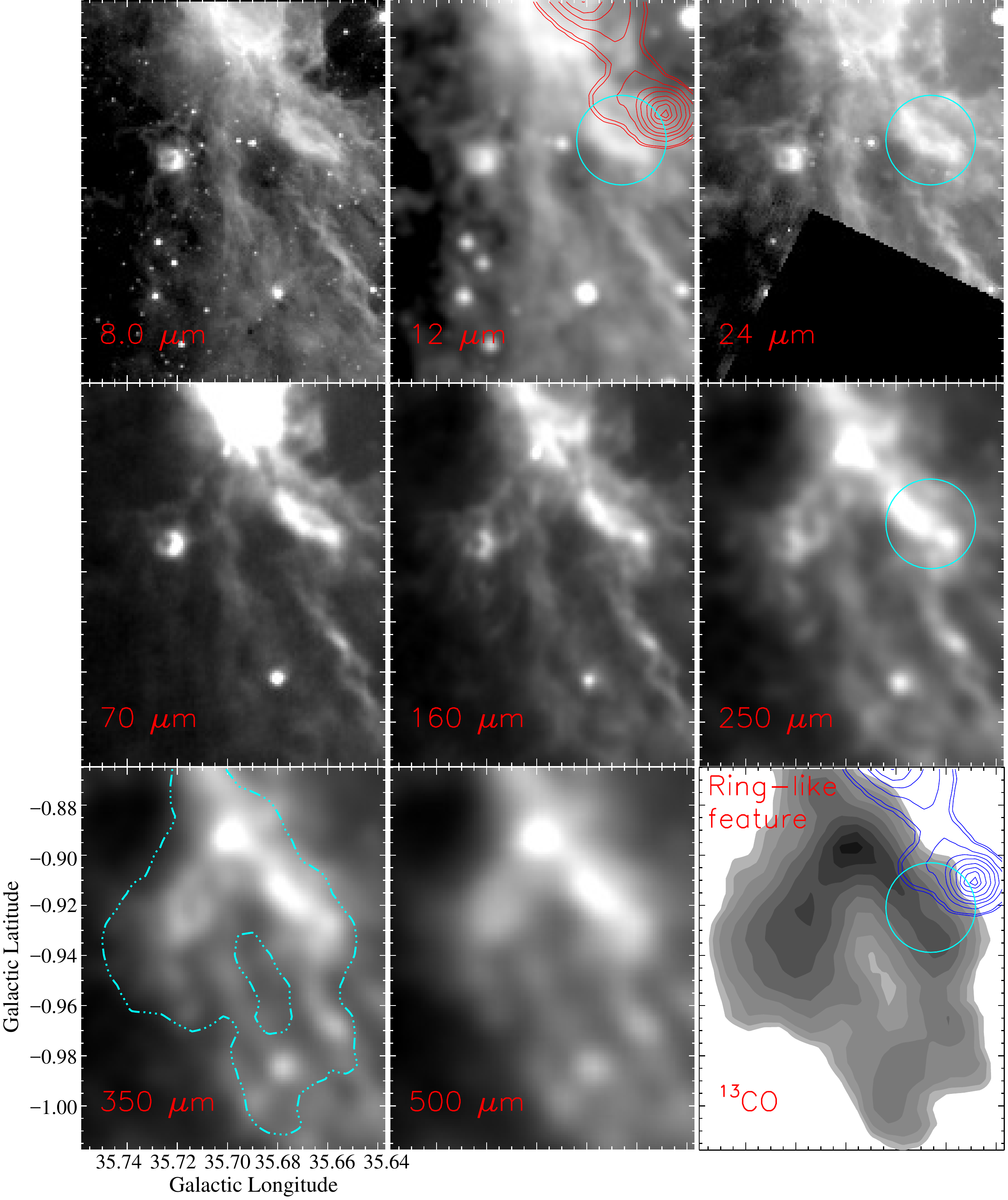}
\caption{\scriptsize A zoomed-in multi-wavelength view of a ring-like feature. 
The panels present images at 8.0 $\mu$m, 12 $\mu$m, 24 $\mu$m, 70 $\mu$m, 160 $\mu$m, 250 $\mu$m, 350 $\mu$m, and 500 $\mu$m, 
and integrated $^{13}$CO contour map, from the GLIMPSE, WISE, MIPSGAL, Hi-GAL, and GRS surveys (from left to right in increasing order). 
The integrated $^{13}$CO contour map is similar to the one shown in Figure~\ref{fg1}a. The integrated $^{13}$CO contour map and the image at 12 $\mu$m are overlaid with the NVSS 1.4 GHz continuum emission contours, which are similar to the one shown in Figure~\ref{fg2}a. 
A $^{13}$CO contour is superimposed on the 350 $\mu$m map with a level of 3.8 K km s$^{-1}$. 
The ring-like feature is prominently evident at wavelengths longer than 160~$\mu$m, and does not contain ionized gas emission at its interior. 
A small feature seen in the infrared and sub-mm images is also highlighted by a circle in four panels.}
\label{fg3}
\end{figure*}

Most of the molecular gas is spatially found in the Galactic southern side of 
the radio source, crs1 (see a yellow box in Figure~\ref{fg2}b). 
A zoomed-in multi-wavelength view of this area (selected field 
$\sim$7$\farcm3$ $\times$ 9$\farcm6$ (or $\sim$7.8 pc $\times$ 10.3 pc)) is presented in Figure~\ref{fg3}. The images are shown at 8.0 $\mu$m, 12 $\mu$m, 24 $\mu$m, 70 $\mu$m, 160 $\mu$m, 250 $\mu$m, 350 $\mu$m, 500 $\mu$m, 
and integrated intensity map of $^{13}$CO (J=1$-$0) from 53 to 62 km s$^{-1}$. 
Note that an almost ring-like feature is evident in the integrated $^{13}$CO intensity map and the {\it Herschel} 
sub-mm images (at 250--500 $\mu$m). 
The central cavity of the molecular emission coincides with a dark region in the infrared images, 
suggesting that there are no stars and dust condensations. 
The integrated $^{13}$CO intensity map and the image at 12 $\mu$m are also superimposed with 
the NVSS 1.4 GHz emission, indicating that the ionized gas emission is not seen inside the ring-like feature.
However, the radio source, crs2 is spatially seen at the most north-western section of the ring-like feature. In the closest vicinity of crs2, within a circle with a radius of 1.25 pc, we find a small feature seen in the infrared and sub-mm images (see a circle in Figure~\ref{fg3} and also the inset in Figure~\ref{fg2}a). 
The feature is resolved in the images below at 100 $\mu$m, while, in the longer wavelength images ($>$ 100 $\mu$m), it becomes a part of the ring-like feature. 
The energetic and feedback of the massive star associated with crs2 seems to influence this feature (see Section~\ref{subsec:feed} for more details). 

The analysis of multi-wavelength images reveals the presence of a semi-ring-like feature (associated with the radio source crs1) and 
an embedded ring-like feature (without detectable ionized gas emission), at the sensitivity of the 
observed NVSS radio map (1$\sigma$ $\sim$0.45 mJy beam$^{-1}$) in the G35.6 site. The presence of a ring-like feature in G35.6 is an excellent 
example for comparing with theoretical models, since observationally it is scarce and 
difficult to find such features with a high degree of geometrical and projectional symmetry. 
Overall, with the help of Figures~\ref{fg2}, and~\ref{fg3}, we have obtained a pictorial multi-wavelength 
view of the G35.6 site.
\subsection{Molecular gas properties in MCG35.6}
\label{sec:coem} 
In this section, we present a kinematic analysis of molecular gas in the MCG35.6.
Note that previously, \citet{paron11} also analyzed the GRS $^{13}$CO data and presented 
the integrated GRS $^{13}$CO intensity map and velocity channel maps mainly toward the G35.6 H\,{\sc ii} 
region (see a small dotted box in Figure~\ref{fg2}b). 
In the present work, we have carefully examined the gas distribution in the direction of our selected target 
field, and have also performed the position-velocity analysis of molecular gas, which was 
lacking in the earlier published work.
\subsubsection{Velocity profiles and molecular cloud}
\label{vvsec:coem} 
\begin{figure*}
\epsscale{1}
\plotone{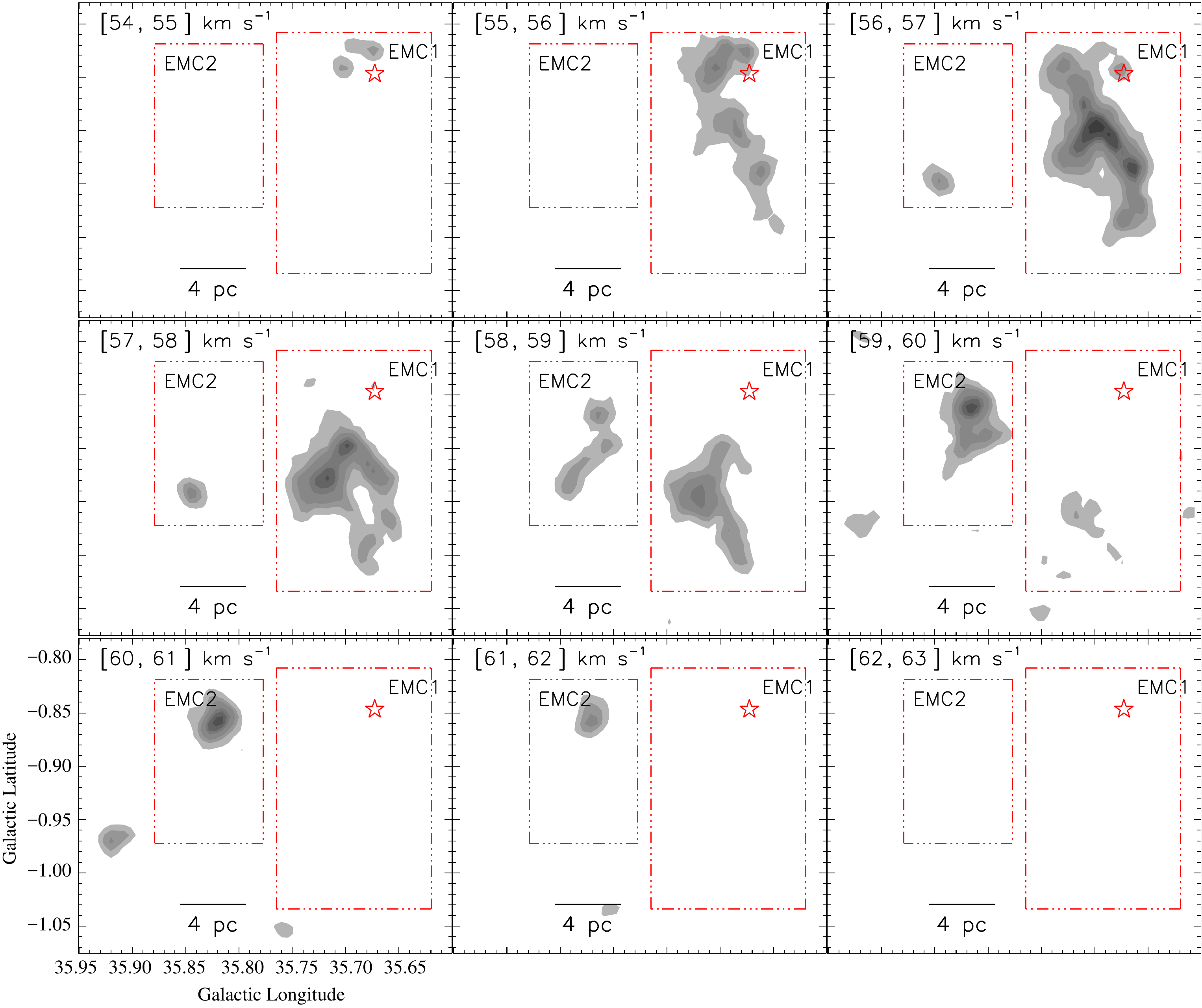}
\caption{\scriptsize The $^{13}$CO (J =1$-$0) velocity channel contour maps in the direction of the site G35.6 (size of the selected field $\sim$21$\arcmin$ $\times$ 17$'$.7; central coordinates: $l$ = 35$\degr$.778; $b$ = $-$0$\degr$.931). The molecular emission is integrated over a velocity interval, which is given in each panel (in km s$^{-1}$). 
The contours are shown with the levels of 1, 2, 3, 5, 6, 7, and 8.5 K km s$^{-1}$. 
Two extended molecular condensations (i.e., EMC1 and EMC2) are labeled in the maps. 
The fields of EMC1 and EMC2 are also shown by dotted-dashed boxes in each map. 
In each panel, a star symbol indicates the position of IRAS 18569+0159. 
The map at [57, 58] km s$^{-1}$ traces an almost ring-like feature.}
\label{fg1x}
\end{figure*}

The integrated GRS $^{13}$CO (J=1$-$0) velocity channel maps (at intervals of 1 km s$^{-1}$) 
are presented in Figure~\ref{fg1x}, which cover the velocity range from 54 to 63 km s$^{-1}$. 
The channel maps reveal two spatially distinct molecular structures, which can also be referred to as extended molecular condensations (EMCs). 
The first one (EMC1) is seen in a velocity range from 55 to 60 km s$^{-1}$, 
and appears to be associated with the radio continuum emission. 
It has an extended morphology showing an almost ring-like feature at a velocity of 57--58 km s$^{-1}$ (see also Figure~\ref{fg3}). 
The second structure (EMC2) appears at velocities from 56 to 62 km s$^{-1}$. 
At low velocities it is compact and peaks at the coordinates {\it l} = 35$\degr$.85, {\it b} = $-$0$\degr$.94, 
while it extends to the north when moving to high velocities having the red-shifted 
peak at the coordinates ({\it l} = 35$\degr$.85, {\it b} = $-$0$\degr$.85). 

\begin{figure*}
\epsscale{0.58}
\plotone{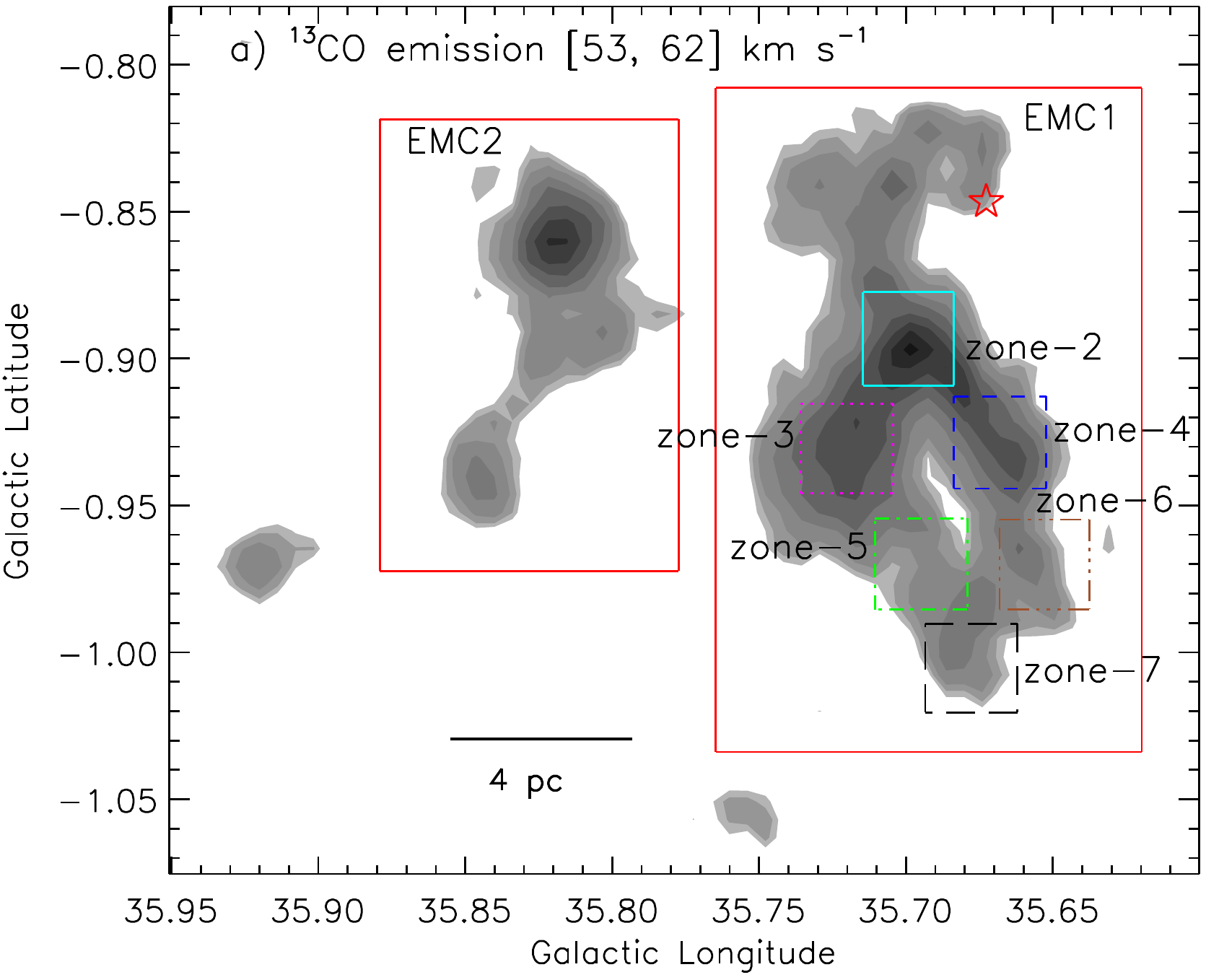}
\epsscale{0.48}
\plotone{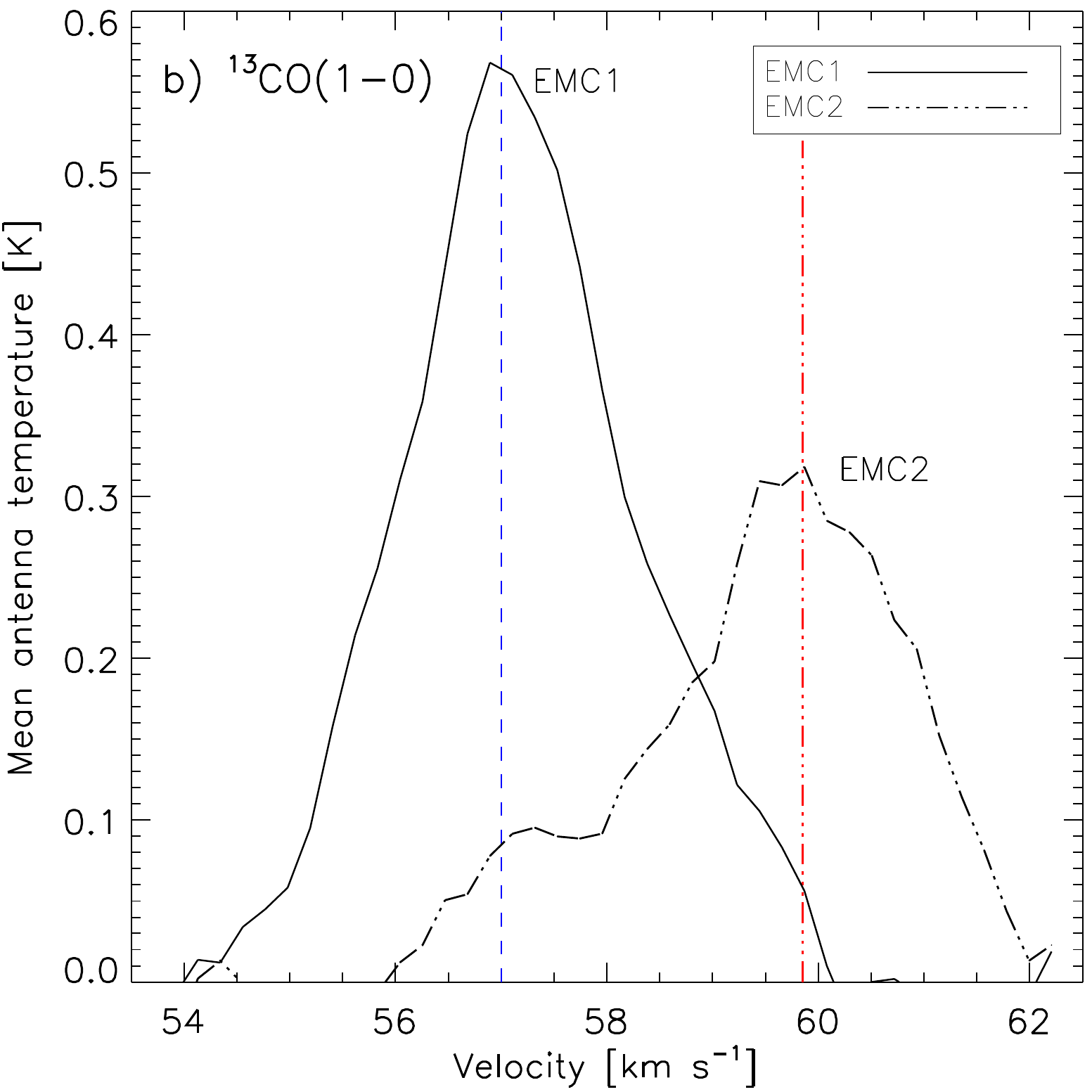}
\epsscale{0.48}
\plotone{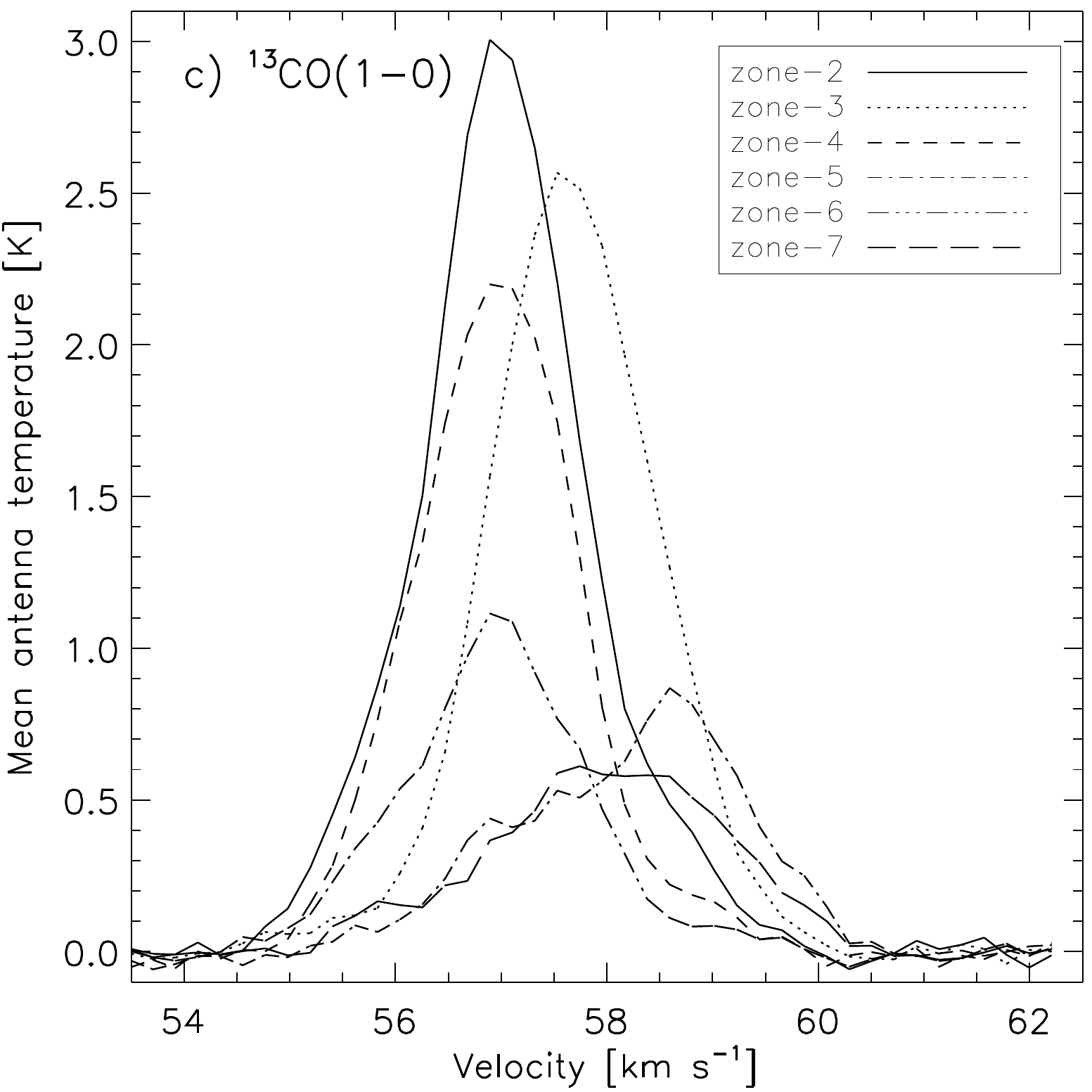}
\caption{\scriptsize a) Molecular emission contour map integrated over a velocity interval from 53 
to 62 km s$^{-1}$. The $^{13}$CO contours are 19.034 K km s$^{-1}$ $\times$ (0.12, 0.15, 0.2, 0.3, 0.4, 0.55, 0.7, 0.85, 0.95). 
Two extended molecular condensations (i.e., EMC1 and EMC2) are labeled in the map.  
In the molecular map, the areas of several fields (i.e. EMC1, EMC2, zone-2, zone-3, zone-4, zone-5, zone-6, and zone-7) are also highlighted by boxes. 
b) The GRS $^{13}$CO(1-0) profile in the direction of two fields, EMC1 and EMC2 (see corresponding box in Figure~\ref{fg2x}a). 
Two velocity peaks are marked by broken lines. 
c) The GRS $^{13}$CO(1-0) spectra in the direction of six small fields linked with the ring-like feature 
(i.e. zone-2 to zone-7; see corresponding boxes in Figure~\ref{fg2x}a). 
In the last two panels, each spectrum is obtained by averaging the area highlighted by a box.}
\label{fg2x}
\end{figure*}

Figure~\ref{fg2x}a shows an integrated intensity map of $^{13}$CO (J = 1-0) from 53 to 62 km s$^{-1}$, 
revealing clearly both the molecular condensations, EMC1 and EMC2 (see also Figure~\ref{fg1x}).  
In the integrated $^{13}$CO map, the ring-like feature is evident at the Galactic southern side of EMC1. 
In Figure~\ref{fg2x}b, we show the observed $^{13}$CO (J=1--0) spectral profiles in the direction of EMC1 and EMC2 
(see respective box in Figure~\ref{fg2x}a).
In the direction of EMC1, the velocity peak is seen at $\sim$57 km s$^{-1}$, while the velocity peak toward EMC2 is 
found at $\sim$60 km s$^{-1}$. These profiles further confirm the presence of 
two molecular cloud components in our selected site. 
Figure~\ref{fg2x}c shows the observed $^{13}$CO (J=1--0) spectra in the direction of six fields 
associated with the ring-like feature (i.e. zone-2 to zone-7; see broken boxes in Figure~\ref{fg2x}a). 
Each spectrum is constructed by averaging the area highlighted by a box in Figure~\ref{fg2x}a. 
In Figure~\ref{fg2x}c, the velocity peaks in the profiles are varying between 57 and 58.5 km s$^{-1}$. 
Based on the comparison of velocity peaks in six profiles, one can find evidence 
of gas motion in the molecular feature. 
One can note that zones 2, 4, and 6 trace the west segment of the ring-like feature, 
while zones 3, 5, and 7 depict the eastern side.
The velocity peaks seen in the direction of zones 2, 4, and 6 have very similar values, 
while the velocity peaks observed in the direction of zones 3, 5, and 7 
have slightly different values, and also have noticeable higher values (i.e. red-shifted) 
than the ones seen toward the zones 2, 4, and 6 (see Figure~\ref{fg2x}c). 
This result also indicates the existence of a velocity spread across 
the ring-like feature (see also Section~\ref{sec:nwcoem}). 

Considering these results and the spatial distribution of molecular gas, 
the ring-like feature has an ellipsoid shape along the line of sight. 
In 3D, we assume that this molecular feature has a spherical oblate morphology.
\subsubsection{Position-velocity diagrams}
\label{sec:nwcoem} 
\begin{figure*}
\epsscale{1}
\plotone{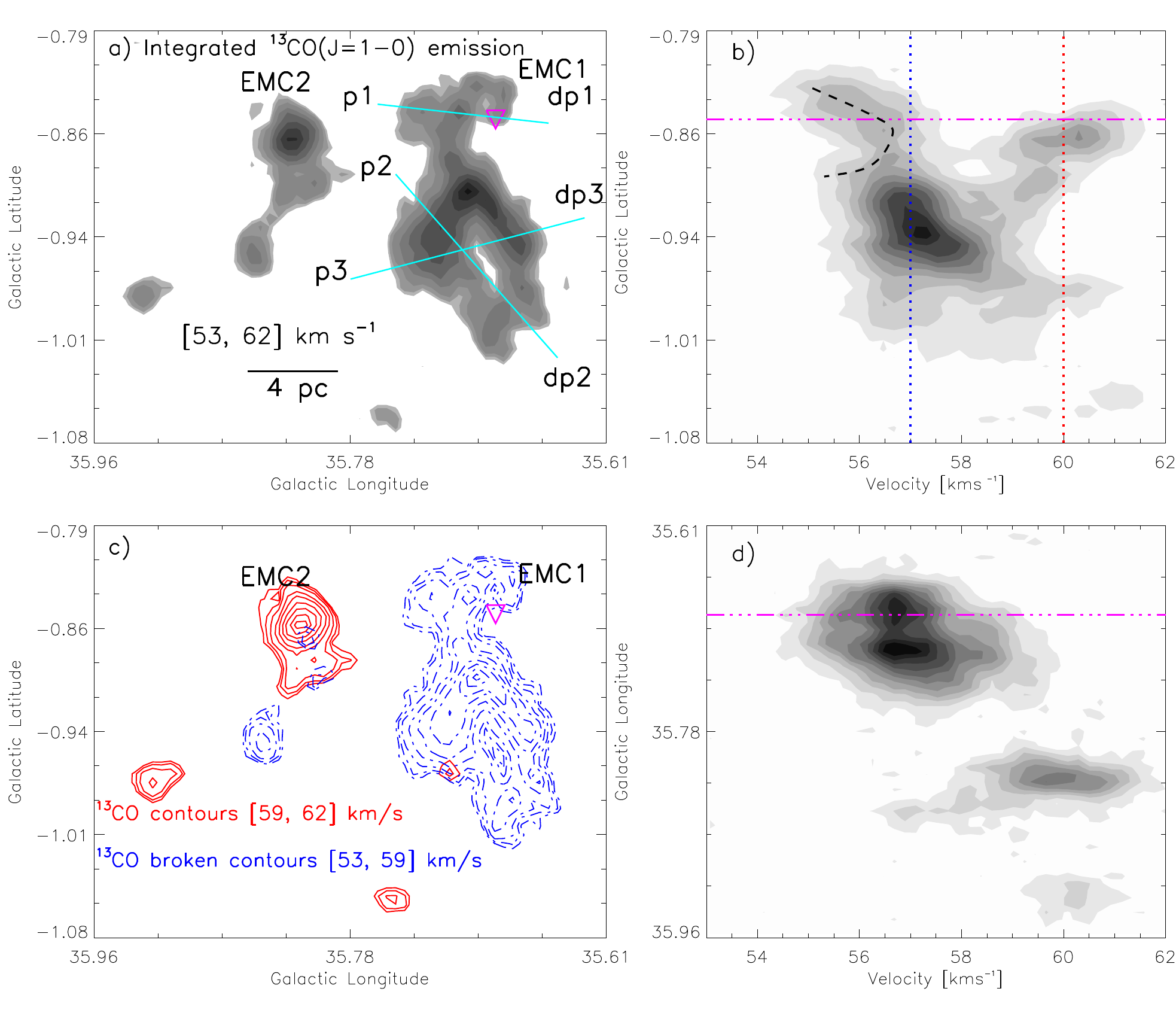}
\epsscale{1.2}
\plotone{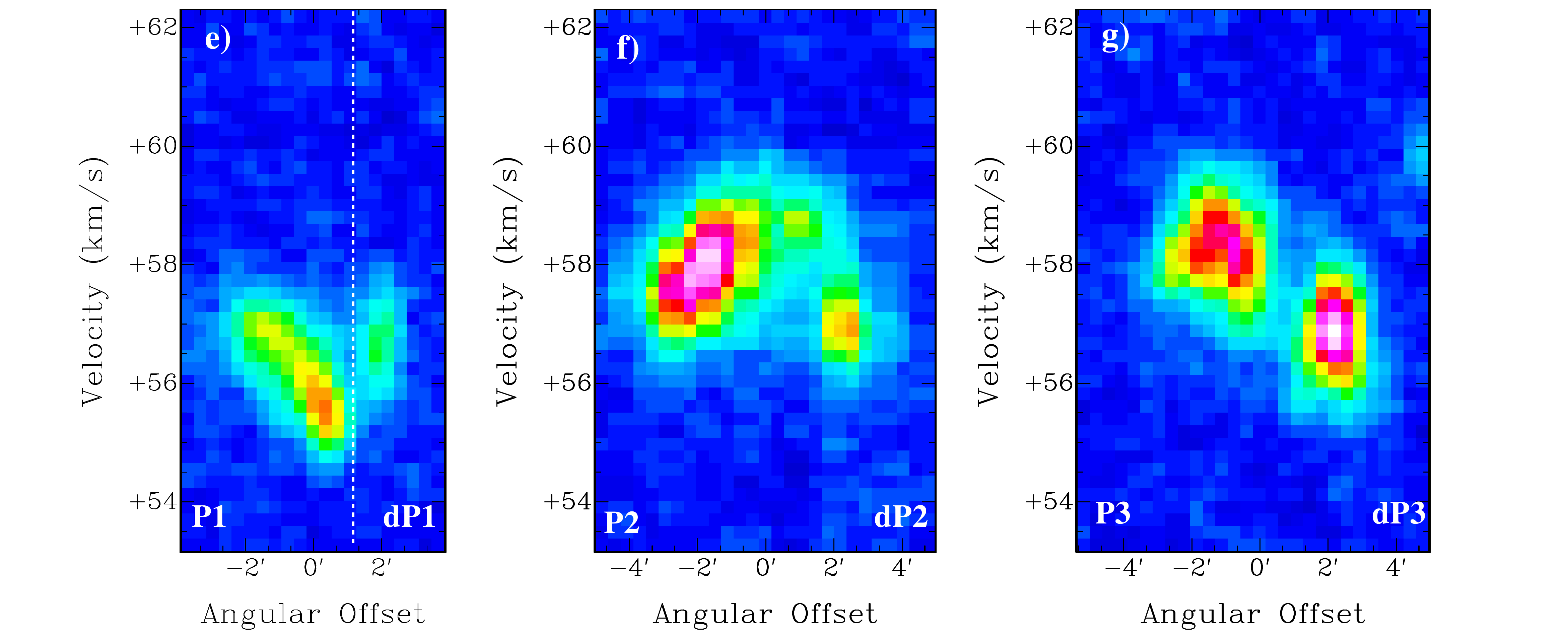}
\caption{\scriptsize a) A contour map of integrated $^{13}$CO emission in the direction of the Galactic H\,{\sc ii} region G35.6. The $^{13}$CO integrated velocity range is from 53 km s$^{-1}$ to 62 km s$^{-1}$. 
The $^{13}$CO contours are similar to those shown in Figure~\ref{fg2x}a. 
Each solid line (in cyan) represents the axis, where the position-velocity diagrams are extracted and are shown in Figures~\ref{fg1}e,~\ref{fg1}f, and~\ref{fg1}g.
b) Latitude-velocity distribution of $^{13}$CO. The $^{13}$CO emission is integrated over the longitude 
from 35$\degr$.61 to 35$\degr$.96. There are two velocity peaks (at $\sim$57 km s$^{-1}$ and $\sim$60 km s$^{-1}$) 
seen in the position-velocity diagram which are highlighted by dotted lines (in red and blue). 
An inverted C-like feature is also marked by a dashed curve (in black) (see also the text).
c) Two molecular components in the direction of the G35.6 site. 
Red contours are the $^{13}$CO emission from 59 to 62 km s$^{-1}$ with levels of 
15.219 K km s$^{-1}$ $\times$ (0.12, 0.15, 0.2, 0.3, 0.4, 0.55, 0.7, 0.85, and 0.95).
Blue contours (dotted-dashed) are the $^{13}$CO emission from 53 to 59 km s$^{-1}$ with levels of 
18.787 K km s$^{-1}$ $\times$ (0.12, 0.15, 0.2, 0.3, 0.4, 0.55, 0.7, 0.85, and 0.95).
d) Longitude-velocity distribution of $^{13}$CO. The $^{13}$CO emission is integrated over the 
latitude from $-$0.$\degr$79 to $-$1.$\degr$08. 
In the first two top left panels, an upside down triangle symbol indicates the position of crs1 and two extended molecular condensations 
(i.e., EMC1 and EMC2) are labeled. In the first two top right panels, a dotted-dashed line shows the position of crs1. 
e) A position-velocity diagram along the axis ``p1--dp1" as shown in Figure~\ref{fg1}a. 
A dotted line indicates the position of crs1. 
f) A position-velocity diagram along the axis ``p2--dp2" as shown in Figure~\ref{fg1}a.
g) A position-velocity diagram along the axis ``p3--dp3" as shown in Figure~\ref{fg1}a.}
\label{fg1}
\end{figure*}

To further study the molecular gas distribution in the direction of our selected target field, 
in Figure~\ref{fg1}, we present the integrated $^{13}$CO intensity map and the position-velocity diagrams.  

In Figure~\ref{fg1}a, we show the molecular $^{13}$CO (J = 1--0) gas emission in the direction 
of the G35.6 site, which is the same as shown in Figure~\ref{fg2x}a. 
In Figures~\ref{fg1}b and~\ref{fg1}d, we present the Galactic position-velocity diagrams of $^{13}$CO emission. 
The $^{13}$CO emission is integrated over the longitude range and latitude range in the 
latitude-velocity diagram (see Figure~\ref{fg1}b) and longitude-velocity diagram (see Figure~\ref{fg1}d), respectively.
The position-velocity diagrams also trace both the molecular components along the line of sight.
A molecular component associated with the G35.6 H\,{\sc ii} region is seen in a velocity range of 53--59 km s$^{-1}$ (i.e. EMC1), 
while the other molecular component is observed at 59--62 km s$^{-1}$ (i.e. EMC2) (see also Section~\ref{vvsec:coem}). 
In Figure~\ref{fg1}c, we present the spatial distribution of molecular 
gas associated with these two molecular components. The figure also reveals that these 
two components (i.e., EMC1 and EMC2) are clearly separated in both space and velocity, and are unrelated molecular clouds.
Hence, our present work is mainly focused toward EMC1. 

Figures~\ref{fg1}e,~\ref{fg1}f, and~\ref{fg1}g show the position-velocity diagrams of $^{13}$CO along the axis ``p1--dp1", ``p2--dp2", and ``p3--dp3", respectively.
The axis ``p1--dp1" is selected toward the radio source crs1, while the other two axes are chosen toward the ring-like feature. 
In the direction of crs1/the semi-ring-like feature, the position-velocity diagrams show hints of an almost inverted C-like structure (see Figure~\ref{fg1}e and also a black dashed curve in Figure~\ref{fg1}b). 
In the velocity space, the detection of an inverted C-like or arc-like 
structure indicates the presence of a bubble/shell \citep{arce11}. 
Hence, in the direction of the semi-ring-like feature, our analysis suggests the presence of an expanding shell with an expansion velocity\footnote[3]{In the velocity space, the expansion velocity corresponds to half of the velocity range for the inverted C-like structure.} of the gas to be $\sim$1.2 km s$^{-1}$. The position of crs1 is highlighted in Figures~\ref{fg1}a--~\ref{fg1}e, 
and nearly appears at the center of the inverted C-like structure (see Figures~\ref{fg1}b and~\ref{fg1}e). 

In Figures~\ref{fg1}b and~\ref{fg1}d, a noticeable velocity spread is also found, where the majority of molecular gas is observed, indicating the gas movement in the ring-like feature (see also Figure~\ref{fg2x}c). In the direction of the ring-like feature, Figures~\ref{fg1}f and~\ref{fg1}g further reveal the observed velocity structures. A velocity spread across the molecular feature is also evident in the diagrams. In the velocity space, two velocity peaks (i.e. 57 and 58 km s$^{-1}$) are seen, likely connected by diffuse gas and therefore suggesting the presence of a single structure with velocity gradients. Higher spectral and angular resolution data can better resolve the structure and the kinematics of the gas in the ring-like structure. As mentioned earlier, there is no ionized gas emission seen inside the ring-like feature, hence the observed velocity spread may not be explained by an H\,{\sc ii} region.
\subsection{Ionized gas properties}
\label{subsec:radio}
In this section, we investigate the spectral types of the powering candidates responsible for the compact radio sources (i.e., crs1 and crs2). 
We have computed the number of Lyman continuum photons ($N_\mathrm{uv}$) for each compact radio source 
using the integrated flux density following the equation of \citet{matsakis76}. 
The ``{\it clumpfind}" IDL program \citep{williams94} can be employed in a given radio continuum map to compute the integrated flux densities. 
Using the 1.4 GHz map, we estimated the integrated flux densities (radii) equal to 160.7 mJy (1.8 pc) and 58.8 mJy (1.25 pc) 
for compact radio sources, crs1 and crs2, respectively. 
The equation of $N_\mathrm{uv}$ is defined by \citet{matsakis76}:
\begin{equation}
\begin{split}
N_{uv} (s^{-1}) = 7.5\, \times\, 10^{46}\, \left(\frac{S_{\nu}}{Jy}\right)\left(\frac{D}{kpc}\right)^{2} 
\left(\frac{T_{e}}{10^{4}K}\right)^{-0.45} \\ \times\, \left(\frac{\nu}{GHz}\right)^{0.1}
\end{split} 
\end{equation}
\noindent where S$_{\nu}$ is the measured total flux density in Jy, D is the distance in kpc, 
T$_{e}$ is the electron temperature, and $\nu$ is the frequency in GHz. 
These calculations were carried out for a distance of 3.7 kpc and for the electron temperature of 10000~K.
Finally, we find $N_\mathrm{uv}$ (or log$N_\mathrm{uv}$) to be $\sim$1.7 $\times$ 10$^{47}$ s$^{-1}$ (47.24) and $\sim$6.3 $\times$ 10$^{46}$ s$^{-1}$ (46.8) 
for compact radio sources, crs1 and crs2, respectively. 
Each compact radio source appears to be associated with a single ionizing star of spectral type B0.5V--B0V 
\citep[see Table II in][and also \citet{smith02}]{panagia73}.  

\citet{paron11} utilized a total flux density of 0.86 Jy at
2.7 GHz for G35.6, and estimated $N_\mathrm{uv}$ to be $\sim$1.0 $\times$ 10$^{48}$ s$^{-1}$. 
The 2.7 GHz continuum data, as part of the Bonn 11 cm survey, were taken with the Effelsburg 100 
meter telescope, and have an angular resolution of 4$'$.3, and 50 mK rms sensitivity \citep{reich84}. 
Note that the 2.7 GHz continuum data have a much coarser resolution compared to the NVSS 1.4 GHz continuum map. 
Hence, the total 2.7 GHz flux density appears to contain radio emission contribution from both the 
radio sources, crs1 and crs2. Using the NVSS 1.4 GHz data, the fluxes of crs1 and crs2 added together result in about 220 mJy. This value is much smaller than the total flux density at 2.7 GHz for G35.6, indicating that the 
Effelsburg map might be detecting some extended emission that is filtered out by the interferometer in the NVSS survey. 
Furthermore, \citet{paron11} used the 1.4 GHz data for the compact radio source, crs2 (or NVSS 185938+020012), 
and computed $N_\mathrm{uv}$ to be $\sim$0.6 $\times$ 10$^{47}$ s$^{-1}$, which is consistent with our estimates. 
Using these values of $N_\mathrm{uv}$, we find at our end that these estimates correspond to a single ionizing 
star of spectral type O9V-O8.5V and B0.5V-B0V for G35.6 and crs2 
(see \citet{martins05}, \citet{panagia73}, and \citet{smith02}, for theoretical values), respectively.   
However, \citet{paron11} reported the spectral type of the ionizing star of G35.6 to be between O7.5V and O9V, considering errors of about ten percent in both the distance and the radio continuum flux at 2.7 GHz. They also reported the spectral type of the ionizing star of crs2 to be later than O9.5V star. 
These authors have not computed the radio spectral type for crs1. 
Based on our independent calculations using the 1.4 GHz data, it appears that the spectral types of 
radio sources reported by \citet{paron11} are overestimated. 
Hence, spectroscopic observations will be needed to confirm the spectral types of the exciting stars. 
Using the photometric data, \citet{paron11} suggested that
a source J18592786+0203057 is the more likely exciting-star candidate of G35.6, while two objects J18593584+0200579 and J18593556+0200488 could be the powering source candidates of crs2 (see Table~1 in their paper). 
Furthermore, in the G35.6 site, a point-like source is also observed in the CORNISH 5~GHz continuum map 
(beam size $\sim$1\farcs5), 
and is referred to as G035.6624-00.8481 in the CORNISH 5 GHz compact source catalog \citep{purcell13}. 
The position of this radio source is indicated by a square in Figure~\ref{fg2}a, and appears about 1$'$.4 
away from the peak position of crs1. 
However, in the direction of crs1, the CORNISH source is seen well within the NVSS radio 
emission contours (see Figure~\ref{fg2}a). 
\citet{purcell13} also computed a total integrated flux density equal to 5.43 mJy for the CORNISH source 
(having an angular size of $\sim$1$'$.5).
With the help of equation~4, we find that the observed 5 GHz flux (having log$N_\mathrm{uv}$ $\sim$45.82 s$^{-1}$) of G035.6624-00.8481 can be 
explained by a single ionizing star of spectral type B1V--B0.5V, which is nearly 
consistent with the spectral type of crs1 estimated using the NVSS 1.4 GHz data.

With the estimates of $N_\mathrm{uv}$ and radii of the H\,{\sc ii} regions (R$_{HII}$), we have also computed the dynamical 
age (t$_{dyn}$) of each compact radio source.
The age of the H\,{\sc ii} region can be obtained at a given radius R$_{HII}$, using the following equation \citep{dyson80}:
\begin{equation}
t_{dyn} = \left(\frac{4\,R_{s}}{7\,c_{s}}\right) \,\left[\left(\frac{R_{HII}}{R_{s}}\right)^{7/4}- 1\right] 
\end{equation}
where c$_{s}$ is the isothermal sound velocity in the ionized gas (c$_{s}$ = 11 km s$^{-1}$; \citet{bisbas09}), 
R$_{HII}$ is previously defined, and R$_{s}$ is the radius of the Str\"{o}mgren sphere (= (3 $N_\mathrm{uv}$/4$\pi n^2_{\rm{0}} \alpha_{B}$)$^{1/3}$, where 
the radiative recombination coefficient $\alpha_{B}$ =  2.6 $\times$ 10$^{-13}$ (10$^{4}$ K/T)$^{0.7}$ cm$^{3}$ s$^{-1}$ \citep{kwan97}, 
$N_\mathrm{uv}$ is defined earlier, and ``n$_{0}$'' is the initial particle number density of the ambient neutral gas. 
Adopting a typical value of n$_{0}$ (=10$^{3}$ cm$^{-3}$), the dynamical ages of the compact radio sources, crs1 
and crs2 are estimated to be $\sim$0.5 and $\sim$0.35 Myr, respectively. 
Note that the H\,{\sc ii} regions are assumed to be uniform and spherically symmetric. 
Recently, \citet{peters10} indicated that the dynamical process of the H\,{\sc ii} regions could be more complex, and dynamical ages can not be estimated accurately with the assumptions of uniform and spherically symmetric. Hence, the derived dynamical ages of the H\,{\sc ii} regions can be considered as the indicative values.

The radio continuum map at 1.4 GHz is also utilized to obtain the electron density ($n_e$) of the ionized gas. 
The formula of ``$n_e$" is presented in \citet{panagia78} under the assumption that the H\,{\sc ii} region has a roughly spherical geometry:
\begin{equation}
\begin{split}
n_{e} (cm^{-3})=3.113\times 10^2 \left(\frac{S_\nu}{\rm{Jy}}\right)^{0.5}\left(\frac{T_e}{{\rm 10^4 K}}\right)^{0.25} 
\left(\frac{D}{{\rm kpc}}\right)^{-0.5} \\ \times\, b(\nu,T)^{-0.5}\theta_R^{-1.5}
\end{split}
\end{equation} 
In the equation above, S$_\nu$, $T_e$, and D are defined as in Equation 4, $\theta_R$ is the angular radius in arcminutes, and 
\[ b(\nu, T)=1+0.3195~{\rm log}\,\left(\frac{T_e}{\rm 10^4 K}\right)-0.2130~{\rm log}\left(\frac{\nu}{\rm 1 GHz}\right).
\] Considering the values of S$_\nu$ and $\theta_R$ for each of the compact radio sources, 
the values of $n_e$ are computed to be 30 and 32 cm$^{-3}$ for crs1 and crs2, respectively. 
Using the CORNISH 5 GHz data, we have also computed the value of $n_e$ to be 9250 cm$^{-3}$ for 
G035.6624-00.8481. Hence, the CORNISH 5 GHz observations suggest the presence of an 
ultracompact H\,{\sc ii} region in the G35.6 site. 
The calculation was carried out for D = 3.7 kpc and T$_{e}$ = 10000 K. 
Note that the electron density of these two radio sources estimated using the 1.4 GHz data appears 
smaller than the one measured assuming optically thin emission. 
Therefore, radio continuum observations with radio interferometers 
at different frequencies can be helpful to better determine the SED of the H\,{\sc ii} regions and obtain a more accurate measurement of the properties of the H\,{\sc ii} regions and the spectral type of the ionizing stars \citep[e.g.][]{panagia75,olnon75,kurtz05,sanchezmonge13}.
\subsection{Feedback effect of massive stars}
\label{subsec:feed}
Our radio data analysis reveals that each compact radio source is excited by a radio spectral type of 
B0.5V star (see Section~\ref{subsec:radio}). 
The radio continuum peak of crs1 is found to be associated with the semi-ring-like structure (see Figure~\ref{fg2}a). It is also observed that the radio source crs2 is 
located toward the north-western part of the ring-like feature (see Figure~\ref{fg3}). 
To explore the feedback of the B0.5V star in its vicinity, we infer the values of three pressure components 
(i.e., pressure of an H\,{\sc ii} region ($P_\mathrm{HII}$), radiation pressure ($P_\mathrm{rad}$), and stellar wind ram 
pressure ($P_\mathrm{wind}$)) driven by a massive star. 
The equations of these pressure components ($P_\mathrm{HII}$, $P_\mathrm{rad}$, and $P_\mathrm{wind}$) are given below \citep[e.g.][]{bressert12}:
\begin{equation}
P_{HII} = \mu m_{H} c_{s}^2\, \left(\sqrt{3N_{uv}\over 4\pi\,\alpha_{B}\, D_{s}^3}\right);\\ 
\end{equation}
\begin{equation}
P_{rad} = L_{bol}/ 4\pi c D_{s}^2; \\ 
\end{equation}
\begin{equation}
P_{wind} = \dot{M}_{w} V_{w} / 4 \pi D_{s}^2; \\
\end{equation}
where $N_\mathrm{uv}$, $c_\mathrm{s}$, and $\alpha_\mathrm{B}$ are previously defined, 
$\mu$= 0.678 \citep[in the ionized gas;][]{bisbas09}, $m_\mathrm{H}$ is the hydrogen atom mass, $\dot{M}_{w}$ is the mass-loss rate, 
V$_{w}$ is the wind velocity of the ionizing source,  $L_\mathrm{bol}$ is the bolometric luminosity of the source, and 
D$_{s}$ is the projected distance from the position of the B0.5V star where the pressure components are evaluated. 
The pressure components driven by massive stars are estimated at angular sizes of the radio sources (i.e. almost twice the radii of crs1 and crs2). 
Using the 1.4 GHz map, we obtained the radii equal to 1.8 pc and 1.25 pc for radio sources, crs1 and crs2, respectively. 
Hence, we have computed the pressure components at D$_{s}$ = 3.5 pc (2.5 pc) for crs1 (crs2).

In the calculations, we have adopted the values of $L_\mathrm{bol}$ = 19950 L$_{\odot}$ \citep{panagia73}, 
$\dot{M}_{w}$ = 2.5 $\times$ 10$^{-9}$ M$_{\odot}$ yr$^{-1}$ 
\citep{oskinova11}, and $V_\mathrm{w}$ = 1000 km s$^{-1}$ \citep{oskinova11}, for a B0.5V star. 
With the help of equations 4, 5, and 6, in the case of crs1 with D$_{s}$ = 3.5 pc, we find
$P_\mathrm{HII}$ $\approx$ 1.5 $\times$ 10$^{-11}$ dynes\, cm$^{-2}$, 
$P_\mathrm{rad}$ $\approx$ 1.7 $\times$ 10$^{-12}$ dynes\, cm$^{-2}$, and
$P_\mathrm{wind}$ $\approx$ 1.1 $\times$ 10$^{-14}$  dynes\, cm$^{-2}$. 
This also gives a total pressure ($P_\mathrm{total}$ = $P_\mathrm{HII}$ + $P_\mathrm{rad}$ + $P_\mathrm{wind}$) driven by a massive star to be $\sim$1.7 $\times$ 10$^{-11}$ dynes\, cm$^{-2}$. 
Similarly, in the case of crs2 with D$_{s}$ = 2.5 pc, we obtain 
$P_\mathrm{HII}$ $\approx$ 1.5 $\times$ 10$^{-11}$ dynes\, cm$^{-2}$, 
$P_\mathrm{rad}$ $\approx$ 3.4 $\times$ 10$^{-12}$ dynes\, cm$^{-2}$, 
$P_\mathrm{wind}$ $\approx$ 2.1 $\times$ 10$^{-14}$  dynes\, cm$^{-2}$, and $P_\mathrm{total}$ $\approx$ 1.8 $\times$ 10$^{-11}$  dynes\, cm$^{-2}$.
These calculations suggest that in both cases, the pressure of the H\,{\sc ii} region is relatively higher than the radiation pressure and the stellar wind pressure. 
Note that in both the cases, $P_\mathrm{total}$ is comparable to the pressure of a typical cool molecular 
cloud ($P_\mathrm{MC}$ $\sim$10$^{-11}$--10$^{-12}$ dynes cm$^{-2}$ for a temperature $\sim$20 K 
and particle density $\sim$10$^{3}$--10$^{4}$ cm$^{-3}$) \citep[see Table 7.3 of][]{dyson80}. 
This implies that the edges of the semi-ring and ring-like features are not wiped out by the impact of the ionized gas. 
Our analysis further suggests that the semi-ring-like morphology can be explained due to the 
ionizing feedback from the B0.5V type star, and the most north-western section of the ring-like feature (see a circle in Figure~\ref{fg3}) is likely to be affected by the radio source crs2. 
Further observations of PDR tracers \citep[e.g.][]{hollenbach97,trevinomorales16,goicoechea17} can better confirm the influence of crs2 on the molecular gas of the ring-like feature.
\subsection{Temperature and column density maps of G35.6}
\label{subsec:temp}
\begin{figure*}
\epsscale{0.46}
\plotone{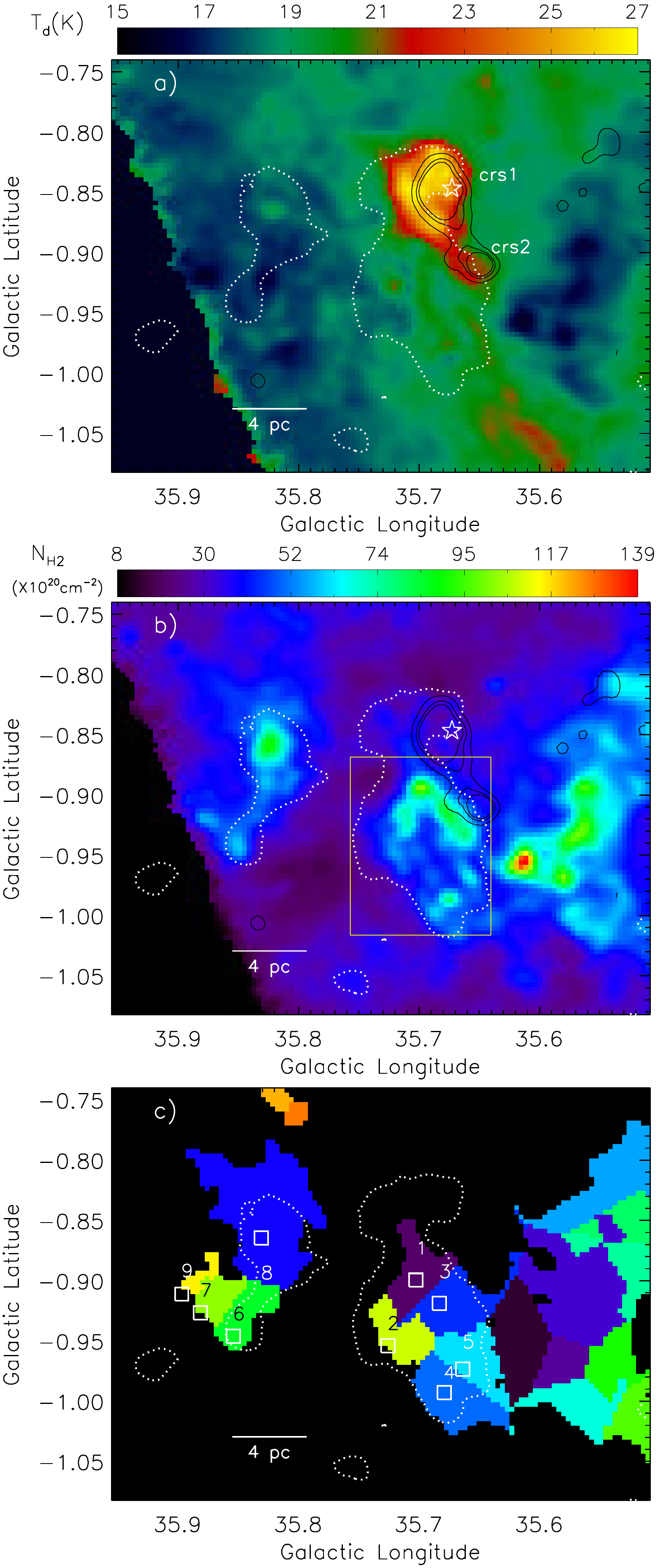}
\caption{\scriptsize a) {\it Herschel} temperature map of the Galactic H\,{\sc ii} region G35.6 
(size of the selected field $\sim27\arcmin  \times 20\farcm5$; central coordinates: $l$ = 35$\degr$.734; $b$ = $-$0$\degr$.914). 
b) {\it Herschel} column density ($N(\mathrm H_2)$) map of the Galactic H\,{\sc ii} region G35.6. 
The extinction can also be inferred using the relation $A_V=1.07 \times 10^{-21}~N(\mathrm H_2)$ \citep{bohlin78}. 
c) The boundary of each identified clump is highlighted along with its corresponding clump 
ID and position (see Table~\ref{tab1}). 
The ring-like feature is also observed in the {\it Herschel} column density map (see a solid yellow box in Figure~\ref{fg4}a). 
In the top two panels, a star symbol indicates the position of IRAS 18569+0159 and 
contours of NVSS 1.4 GHz continuum emission (solid black) are superimposed with 8, 10, and 60\% of 
the peak value (i.e., 31.7 mJy/beam). 
In all the panels, the $^{13}$CO emission contour (dotted white) is also shown from 53 to 62 km s$^{-1}$ 
with a level of 2.15 K km s$^{-1}$. 
In each panel, the scale bar at the bottom-left corner corresponds to 4 pc (at a distance of 3.7 kpc).}
\label{fg4}
\end{figure*}

In this section, to probe the embedded structure and dust condensations, we present the {\it Herschel} temperature and column density maps (resolution $\sim$37$\arcsec$) of the MCG35.6. 
The temperature and column density maps are shown in Figures~\ref{fg4}a and~\ref{fg4}b, respectively. 
The procedures for producing the {\it Herschel} temperature and column density maps were described in Section~\ref{sec:herd}.

In the {\it Herschel} temperature map, the compact radio continuum sources (i.e., crs1 and crs2) 
are seen associated with warmer emission ($T_\mathrm{d}$ $\sim$22-27 K; 
see Figure~\ref{fg4}a), confirming the influence of the radio continuum sources 
on the dense gas as discussed in Section~\ref{subsec:feed}. 
In the {\it Herschel} column density map, several condensations are observed in the MCG35.6 (see Figure~\ref{fg4}b).
The ``{\it clumpfind}" IDL program is executed to identify the clumps and to estimate their total column densities. 
Several column density contour levels were used as an input parameter for the ``clumpfind", and the lowest contour level was considered at 3$\sigma$. Nine clumps are found in the MCG35.6 and are highlighted in Figure~\ref{fg4}c. 
Furthermore, the boundary of each clump is also presented in Figure~\ref{fg4}c. 
With the knowledge of the total column density of each clump, we have also determined 
the mass of each {\it Herschel} clump using the following equation:
\begin{equation}
M_{clump} = \mu_{H_2} m_H Area_{pix} \Sigma N(H_2)
\end{equation}
where $\mu_{H_2}$ is assumed to be 2.8, $Area_{pix}$ is the area subtended by one pixel, and 
$\Sigma N(\mathrm H_2)$ is the total column density. 
The mass of each {\it Herschel} clump is provided in Table~\ref{tab1}. 
The table also gives an effective radius and a peak column density of each clump, which are extracted from the {\it clumpfind} algorithm. 
The clump masses vary between 175 M$_{\odot}$ and 3390 M$_{\odot}$. 
The peak column densities of these clumps vary between $\sim$4.2 $\times$ 10$^{21}$ cm$^{-2}$ (A$_{V}$ $\sim$4.5 mag) and 
$\sim$10 $\times$ 10$^{21}$ cm$^{-2}$ (A$_{V}$ $\sim$10.5 mag) (see Table~\ref{tab1} and also Figure~\ref{fg3}b). 
Here, we used a relation between optical extinction and hydrogen column density \citep[i.e. $A_V=1.07 \times 10^{-21}~N(\mathrm H_2)$;][]{bohlin78}. 
We find that the EMC1 contains five {\it Herschel} clumps (i.e., nos. 1, 2, 3, 4, and 5) and the total mass of these five clumps is $\sim$5535 M$_{\odot}$.
Interestingly, the ring-like feature embedded within EMC1 is also traced in the {\it Herschel} column density map (see a solid yellow box in Figure~\ref{fg4}b). 
Furthermore, at least five clumps (i.e., nos. 1, 2, 3, 4, and 5) appear to be nearly regularly spaced along the ring-like feature. 
The mean separation between the clumps is computed to be 
164$''$.6$\pm$66$''$.5 (or 2.95$\pm$1.20 pc).
One can also find at least three {\it Herschel} clumps (i.e., nos. 6, 7, and 8) toward EMC2. 
\subsection{Polarimetric properties of MCG35.6}
\label{subsec:polu}
In this section, we analyze the background starlight polarization from the GPIPS and the dust polarized emission from the {\it Planck}, to study the magnetic field morphology and properties. 
The POS magnetic field direction can be inferred through the polarization observations from and due to dust grains. 
Polarization of radiation emitted by spinning dust grains is parallel to their long axis and 
the projected field direction can be obtained through the polarization vectors rotated by 90 degrees. 
The physical process (i.e. radiative torque mechanism) concerning the alignment of dust gains and magnetic field is described in \citet{dolginov76} \citep[see also][]{lazarian07,andersson15}. Using the polarization vectors of background stars, one can also obtain the field direction in the POS parallel to the direction of
polarization \citep{davis51}. Polarization of background starlight can be explained due to the dichroism by non-spherical dust grains.
\subsubsection{Magnetic field structure}
\label{vsec:plmg}
\begin{figure*}
\epsscale{0.73}
\plotone{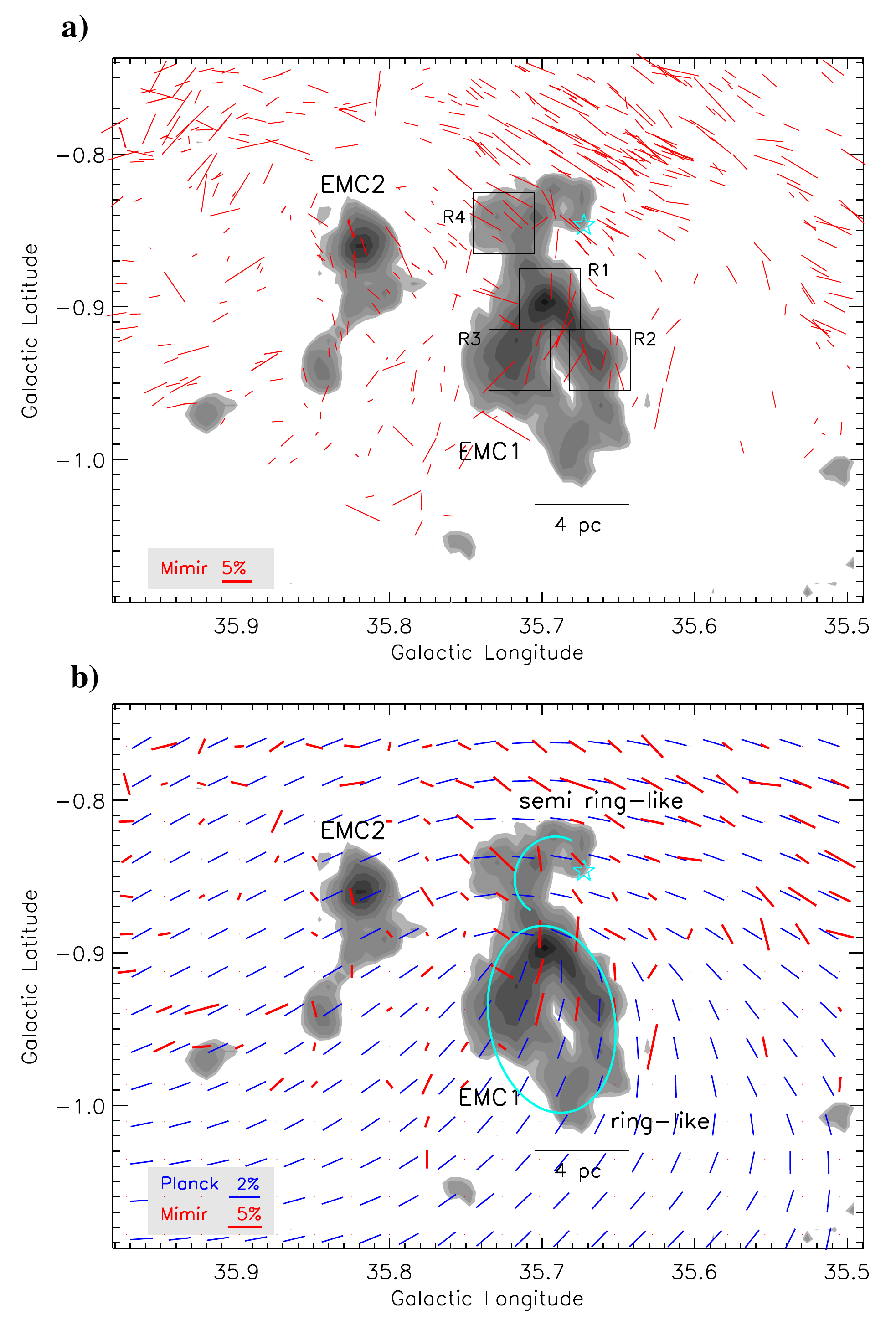}
\caption{\scriptsize The  polarization measurements toward the MCG35.6. 
a) The GPIPS H-band polarization vectors (in red color) of high quality background stars are drawn on the molecular intensity map. 
These stars are selected using the criteria {\footnotesize$P/\sigma_{P}\geq2$, $\sigma_{P}\le5$, $H \leq 13$ and $J-H > 1$}. 
The four subregions targeting the dense clumps/cores are marked by square boxes. 
The length of each vector shows the degree of polarization. 
The orientation of each vector shows the Galactic position angle of polarization. 
b) The {\it Planck} polarization vectors (in blue) from the dust emission at 353 GHz are overlaid on the molecular intensity map. 
The vectors are rotated by 90$^\circ$ to infer the POS magnetic field directions. 
The mean GPIPS polarization vectors (in red) are also overplotted on the molecular map and are obtained 
by spatially averaging the values matching the same pixel scale of the {\it Planck} data. 
In both the panels, the background map is similar to the one shown in Figure~\ref{fg1}a. 
In each panel, a reference vector in the lower left corner is also shown for comparison.}
\label{fg5}
\end{figure*}

To probe the magnetic field morphology using starlight polarimetry, it is important to examine the 
relative polarization position angle orientations of the background stars with respect to the molecular cloud. 
In Section~\ref{sec:obser}, we mentioned about the selection conditions to obtain the sources with reliable polarimetric information.
These selected sources were further exposed to a color condition of J$-$H $>$ 1 to depict the background stars that can 
successfully trace the POS magnetic field direction. 
A total of 462 stars have been selected in our target region around the G35.6 site. 
Figure~\ref{fg5}a shows the observed H-band polarization of background starlight overlaid on the integrated molecular map.
In Figure~\ref{fg5}a, the degree of polarization is depicted by the length of a vector, whereas the angle of a vector shows 
the polarization Galactic position angle. 
In Figure \ref{fg5}a, in EMC1, we have also highlighted four subregions (i.e. R1, R2, R3, and R4) 
by boxes and each box has a size of $\sim$2$'$.4 $\times$ 2$'$.4.
These subregions are associated with the areas of high column density in the {\it Herschel} column density map (see also Figure~\ref{fg4}b).

In Figure \ref{fg5}b, we show the {\it Planck} polarization vectors (in blue) from the dust emission at $353\,\mathrm{GHz}$. 
The vectors are rotated by 90$^\circ$ to show the POS magnetic field direction. The degree of polarization from the dust emission 
observed in the POS also depends on the variations of magnetic field along the line-of-sight (LOS). 
This causes de-polarization effects in the measurement. Since we are studying only 
the magnetic field direction using the {\it Planck} data, we average all the polarization values 
across the region, resulting in equal values in the degree of polarization. 
This exercise leads to the same length for all the vectors, keeping their original position angles. 
The {\it Planck} data at 353 GHz have a beam size of 5$'$ with a pixel scale of 1$'$.5 ($\sim$1.6 pc at a distance of 3.7 kpc). 
In order to compare the magnetic field direction from the 353 GHz dust emission polarimetry and the NIR polarimetry, we calculate the mean polarization 
of the GPIPS data by spatially averaging the values matching the same pixel scale of the {\it Planck} data. 
The GPIPS polarization data are gridded into maps with box regions of 1$'$.5 $\times$ 1$'$.5. All the Stokes $Q$ and $U$ 
values within each box region are combined by variance-weighted mean with their corresponding errors. 
The mean stokes values for each region are then used to compute the Ricean corrected de-biased polarization 
percentages and position angles for each box region \citep[e.g.][]{clemens12a}. The mean GPIPS polarization values are shown in red, along with the {\it Planck} polarization vectors (in blue) in Figure \ref{fg5}b. 

In Figure \ref{fg5}b, the {\it Planck} and GPIPS polarization vectors are in agreement with their relative position angles. 
However, one can notice differences in the orientation of the {\it Planck} and GPIPS vectors toward the molecular condensations (or areas of high column density).
The {\it Planck} data show higher SNR results toward the molecular condensations due to the strong dust emission, whereas the GPIPS data 
toward the molecular condensations have a few measurements with low SNR due to high extinction. 
In EMC1, the distribution of polarization vectors appears different toward the ring-like feature and the semi-ring-like structure, implying a change in the magnetic field morphology between the two features. 

\begin{figure*}
\epsscale{0.48}
\plotone{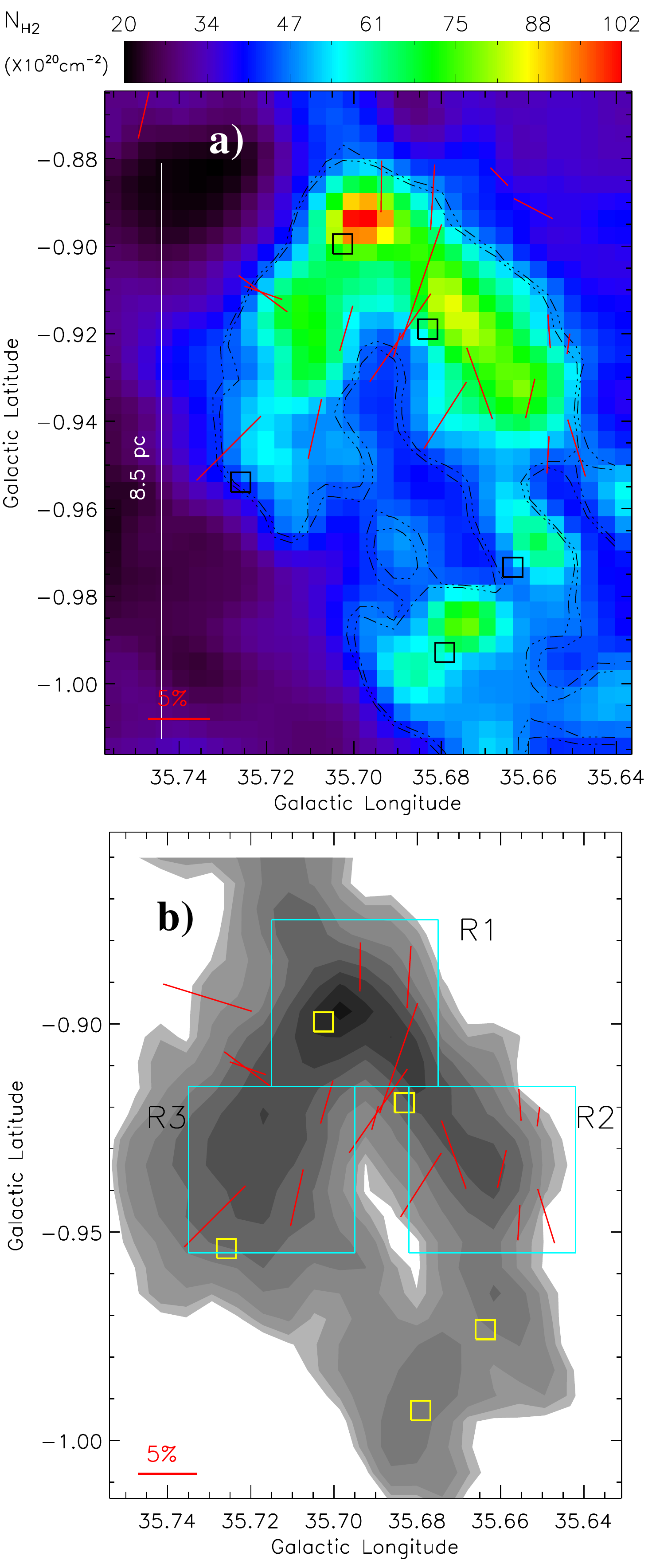}
\caption{\scriptsize The H-band polarimetric measurements toward the ring-like feature (see Table~\ref{tab2}). 
a) Overlay of GPIPS H-band polarization vectors (in red color) of high quality background stars on the {\it Herschel} column density map. 
Column density contours are also shown with levels of [4.5, 4.67] $\times$ 10$^{21}$ cm$^{-2}$.
b) Overlay of GPIPS H-band polarization vectors (in red color) of high quality background stars on the integrated $^{13}$CO emission map. 
Three subregions (i.e. R1, R2, and R3) targeting the areas of high column density are marked in big square boxes. 
A reference vector is also shown in each panel. In both the panels, the {\it Herschel} clumps are also marked by small squares. 
At least five clumps appear to be distributed in an almost regularly spaced manner along the ring-like feature.}
\label{fg5x}
\end{figure*}

Figures~\ref{fg5x}a and~\ref{fg5x}b show a zoomed in view of the ring-like feature using the {\it Herschel} column density and integrated molecular maps, respectively.
These maps are also overlaid with the GPIPS H-band polarization 
vectors (having {\footnotesize$P/\sigma_{P}\geq2$, $\sigma_{P}\le5$, $H \leq 13$ and $J-H > 1$}) and the {\it Herschel} clumps.
A total of 17 stars are detected toward the ring-like feature, tracing the POS magnetic field direction. 
Table~\ref{tab2} lists the 2MASS photometric magnitudes, degree of polarization, and polarization Galactic position angles of these 17 stars. 
The degree of polarization for the 17 sources varies between $\sim$2 to 6\%,  whereas the polarization Galactic position angles have 
a dispersion of around 25$^\circ$ (see Table~\ref{tab2}). 
With the help of the molecular gas distribution and the polarization data, we can characterize the orientation of polarization position angles with respect to the semi-major axis of the molecular structure. Fitting an ellipse to the ring-like feature (see Figure~\ref{fg5x}b), we find that the angle of the major axis of the ellipse is about 10$\degr$. The mean polarization angles in three subregions R1, R2, and R3 are computed to be about 175$\degr$, 177$\degr$, and 169$\degr$, respectively, and are measured from North-up, counter-clockwise (see Figure~\ref{fg5}a). The difference between the polarization angle and the ellipse major axis is about 15$\degr$. This indicates that the POS magnetic field is nearly parallel to the major axis of the ring-like feature. 
In Figures~\ref{fg5}a and~\ref{fg5}b, the semi-ring-like feature is seen at the northern part of EMC1. 
Similarly, we fit a semi-ellipse to the semi-ring-like feature and obtain the angle of the major axis to be 150$\degr$. 
The mean polarization angle in the subregion R4 linked with the semi-ring-like feature is about 55$\degr$. 
The difference between the polarization angle and the major axis is about 95$\degr$, indicating 
that the POS field orientation is perpendicular to the molecular structure associated with the subregion R4.
\subsubsection{Magnetic field strength} 
\label{subsec:polmag}
The POS magnetic field strength can be computed using the observed polarimetric data, 
in combination with gas motions and density from the molecular line data, 
following the equation given in \citet{chandrasekhar53} (hereafter CF-method). The CF-method as modified 
by \citet{ostriker01} can be presented in the following manner: 
\begin{equation}
B_{pos} = 0.5 \times \Big{(} \frac{4}{3} \: \pi \: \rho \Big{)}^{0.5}  \times \: \frac{\sigma_{v}}{\alpha}
\end{equation}
where $B_{pos}$ is the POS magnetic field strength (in $\mu$G), $\alpha$ is the polarization position angle (P.A.) dispersion (in radians), 
$\rho$ is the gas volume mass density (in g cm$^{-3}$), and $\sigma_{v}$ is the $^{13}$CO gas velocity dispersion (in cm s$^{-1}$).
\citet{ostriker01} also suggested that the modified CF-method is applicable only when P.A. dispersion values are within $25^\circ$ (i.e. $\alpha \leq 25^\circ$). 

To measure the magnetic field strength in the denser subregions within EMC1, we have carefully chosen four subregions (R1-R4) 
 and the size of each subregion is 2$'$.4 $\times$ 2$'$.4 (as highlighted by boxes in Figure \ref{fg5}a). 
 These subregions are centered around the {\it Herschel} clumps/molecular condensations.
They are small enough to probe the magnetic field strength to few parsec scales and also big enough to have sufficient high 
quality polarization stars to measure the P.A. dispersion accurately. 

The P.A. dispersion ($\alpha$) is calculated by finding the standard deviation of the P.A. values within each box region. 
Since the P.A. values have an ambiguity of 0$^\circ$ and 180$^\circ$, dealiasing is necessary. 
This is done by finding the minimum of the dispersion values obtained for different sets of biased P.A. values. 
The biasing is carried out by stepping the P.A. values through increments of 1$^\circ$ between 0$^\circ$ and 180$^\circ$ 
producing different sets of biased P.A. values. The standard deviation for each set is computed and the minimum of the 
different standard deviation values is chosen as the final dispersion value. 
Note that typically, five polarization vectors are needed to obtain an accurate 
value of the P.A. dispersion. 
In the present case, we have only three polarization vectors in the subregion R3, where the 
derived P.A. dispersion is about 17$\degr$. Considering this dispersion value, 
we can obtain a lower limit of the B-field in the subregion R3. 

\begin{figure*}
\epsscale{1}
\plotone{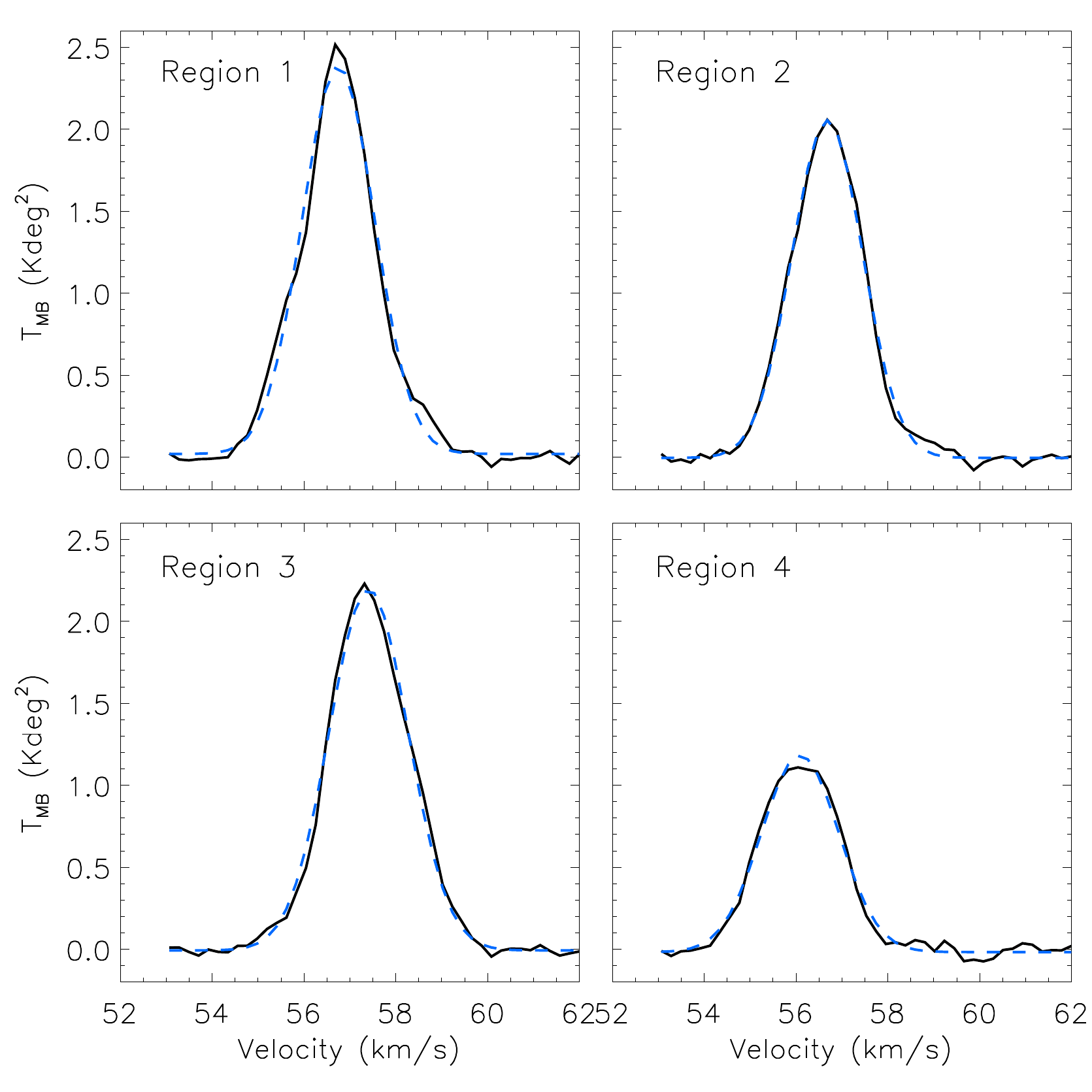}
\caption{\scriptsize A mean spectrum toward each subregion (see also Figure~\ref{fg5}a). 
In each subregion, the observed mean profile is shown by a solid curve, 
while the dashed curve indicates the resulting Gaussian fit.}
\label{cfx}
\end{figure*}

The gas velocity dispersion ($\sigma_v$) is computed from the $^{13}$CO line profile for each pixel within the subregion. 
The line profile is fitted with a Gaussian to obtain the velocity dispersion as the width of the Gaussian. 
Multiple dispersion values within our selected box regions are then combined by median filtered mean to obtain the 
average dispersion value to be used in the CF-method for each region. 
A mean spectrum toward each subregion is presented in Figure~\ref{cfx}.

The {\it Herschel} column density map is used to determine the volume densities for the four subregions in EMC1. 
Based on the ring-like structure assuming a face-on view, we assume a spherically oblate morphology for the cloud structure. 
Using this morphology we estimate the cloud thickness to be $\sim$4 pc. 
Based on the cloud thickness and the column densities, we calculate the volume number density of the region. 
The volume densities were gridded as a map having a spatial resolution of 14$\arcsec$. 
All the volume densities within our each selected box region are then averaged to obtain the mean volume density. 
This is then converted into volume mass density $\rho$ by taking into account the 
molecular hydrogen's weight (2 $\times$ 1.00794 $\times$ 1.67 $\times$ 10$^{-24}$ gm) and a factor of 1.36 to account for helium 
and heavier elements. 
With the application of the CF-method, the POS magnetic field strength in each subregion has been estimated (see Table~\ref{tab3}). 
Using the values of $B_\mathrm{pos}$, we have also estimated magnetic pressure, $P_{mag}$ (= $B_{pos}^2(G)/8\pi$; dynes cm$^{-2}$), equal to 
$\sim$1.8 $\times$ 10$^{-11}$, $\sim$4.8 $\times$ 10$^{-12}$, $\sim$4.8 $\times$ 10$^{-12}$, and 
$\sim$1.6 $\times$ 10$^{-11}$ dynes cm$^{-2}$ for the subregions R1, R2, R3, and R4, respectively.
The average magnetic field strength for all the subregions is computed to be about $15\,\mathrm{\mu G}$ and the 
corresponding magnetic pressure is estimated to be $\sim$8.9 $\times$ 10$^{-12}$ dynes cm$^{-2}$.  

We have also obtained the total field strength by assuming equipartition of magnetic, gravitational, and kinetic energy using a relation given in \citet{lada04}
(i.e. $B_\mathrm{EP}$ = $\Delta$V$_{NT}$ $\times$ (3$\pi$$\rho$/2{\it ln}2)$^{1/2}$; where $\Delta$V$_{NT}$ is the nonthermal component of the observed line
width). The non-thermal velocity dispersion is given by:
\begin{equation}
\Delta V_{\rm NT} = \sqrt{\frac{\Delta V^2}{8\ln 2}-\frac{k T}{29 m_H}} = \sqrt{\frac{\Delta V^2}{8\ln 2}-\Delta V_{\rm T}^{2}} ,
\label{sigmanonthermal}
\end{equation}
where $\Delta V$ (=(8\,{\it ln}\,2\,$\sigma _{v}$$^{2}$)$^{1/2}$) is the measured full width at half maximum 
(FWHM) of the observed $^{13}$CO spectra, 
$\Delta V_{\rm T}$ (= $(k T/29 m_H)^{1/2}$) is the thermal broadening for $^{13}$CO at T, 
and $m_H$ is the mass of the hydrogen atom. 
In the {\it Herschel} temperature map, the ring-like feature is depicted in a temperature range of about 18--20 K. 
Using the values of $\Delta V$ and $\Delta V_{\rm T}$ (at T = 20 K; see Section~\ref{subsec:temp}), we derive $\Delta$V$_{NT}$, 
and then the values of B$_{EP}$ are estimated to be 9.5, 8.8, 8.8, and 6.3 ${\mu G}$ for subregions R1, R2, R3, and R4, respectively.

The values of B$_{EP}$ are slightly lower than that of $B_{pos}$. 
In general, an equipartition stage indicates that all the forces are equal. 
However, based on the observations, we find that the magnetic field may have played a dominant role in the cloud formation and its evolution. 
\subsubsection{Mass to Flux Ratio}
\label{subsec:mfrt}
The knowledge of the magnetic field strength and ``mass-to-flux (M/$\Phi$)" ratio 
can enable to infer the magnetically subcritical and supercritical clouds/cores. These parameters can also allow to observationally 
assess the molecular clouds/cores stability and can help to understand the ongoing physical process in the cloud.
However, the observational measurement of M/$\Phi$ of a cloud core is a difficult experiment \citep[e.g.][and references therein]{clemens16}. 

The normalized mass-to-flux ratio ($\overline{M/\Phi_{B}}$) can be calculated from magnetic field strength as follows \citep{crutcher04}:

\begin{equation}
 \overline{M/\Phi_{B}} = 7.6 \times 10^{-21} \frac{N_{H_{2}}}{|B|}
\end{equation}
where $N_{H_{2}}$ is the column density (in cm$^{-2}$) and $|B|$ is the 3-Dimensional magnetic field strength (in $\mu$G).
Note that in equation 13, $|B|$ refers to the 3D magnetic field strength while our estimates are the POS magnetic field strength. 
Assuming that the LOS magnetic field is zero then one can replace $|B|$ with $B_\mathrm{pos}$. 
However, in this case, $B_\mathrm{pos}$ will represent a lower limit to the total field strength. 
Hence, the mass-to-flux ratios of the four subregions are upper limit values. 
To obtain the corrected normalized mass-to-flux ratio, \citet{planck16d} report a correction factor 
of 1/3 and 3/4 for cloud geometries having magnetic field perpendicular and parallel to their major axis, respectively.
In the ring-like feature, the magnetic field is parallel to its major axis.
Hence, we applied a correction factor of 3/4 to the mass-to-flux ratio calculation 
for subregions R1, R2, and R3. In the semi-ring-like feature, the magnetic field is 
perpendicular to its major axis, favouring a correction factor of 1/3 for computing the mass-to-flux ratio in the subregion R4. 

Using the POS magnetic field strength and the {\it Herschel} column density map, 
we have calculated the normalized mass-to-flux ratio for four subregions in EMC1. 
Three subregions (R1-R3) in EMC1 have mass-to-flux ratio greater than 1, implying 
the cloud is supercritical, 
while the subregion R4 is subcritical due to mass-to-flux ratio less than 1. 
In the subcritical region, the magnetic field dominates over gravity.
On the other hand, in the supercritical region, the magnetic field cannot prevent gravitational collapse and hence 
the cloud can evolve and fragment into denser cores. 
Table~\ref{tab3} lists the magnetic properties for all the four subregions (R1-R4) in EMC1.
\subsection{Young stellar population in MCG35.6}
\label{subsec:phot2}
\subsubsection{Selection of YSOs}
\label{subsec:phot1}
In this section, to identify embedded YSOs within the MCG35.6, we have utilized the photometric data at 1--24 $\mu$m 
extracted from various infrared photometric surveys (e.g., MIPSGAL, GLIMPSE, UKIDSS-GPS, and 2MASS). 
These data sets enable us to infer infrared excess sources. 
The infrared excess emission around YSOs is explained by the presence of an envelope and/or circumstellar disk. 
Previously, \citet{paron11} also presented 36 YSOs identified using the 2MASS and GLIMPSE 3.6--8.0 $\mu$m 
photometric data sets, which were restricted only within the EMC1 \citep[see Figure~6 and Table~2 in][]{paron11}. 
In the present work, we have utilized the UKIDSS-GPS NIR data, in combination with the GLIMPSE 3.6--8.0 $\mu$m and 
MIPSGAL 24 $\mu$m data for depicting infrared excess sources. 
Note that the UKIDSS-GPS NIR survey is three magnitudes deeper than 2MASS. 
Hence, our present analysis will allow us to trace more deeply embedded and faint YSOs compared to previously published work. 
A brief description of the selection of YSOs is as follows.\\

\begin{figure*}
\epsscale{0.90}
\plotone{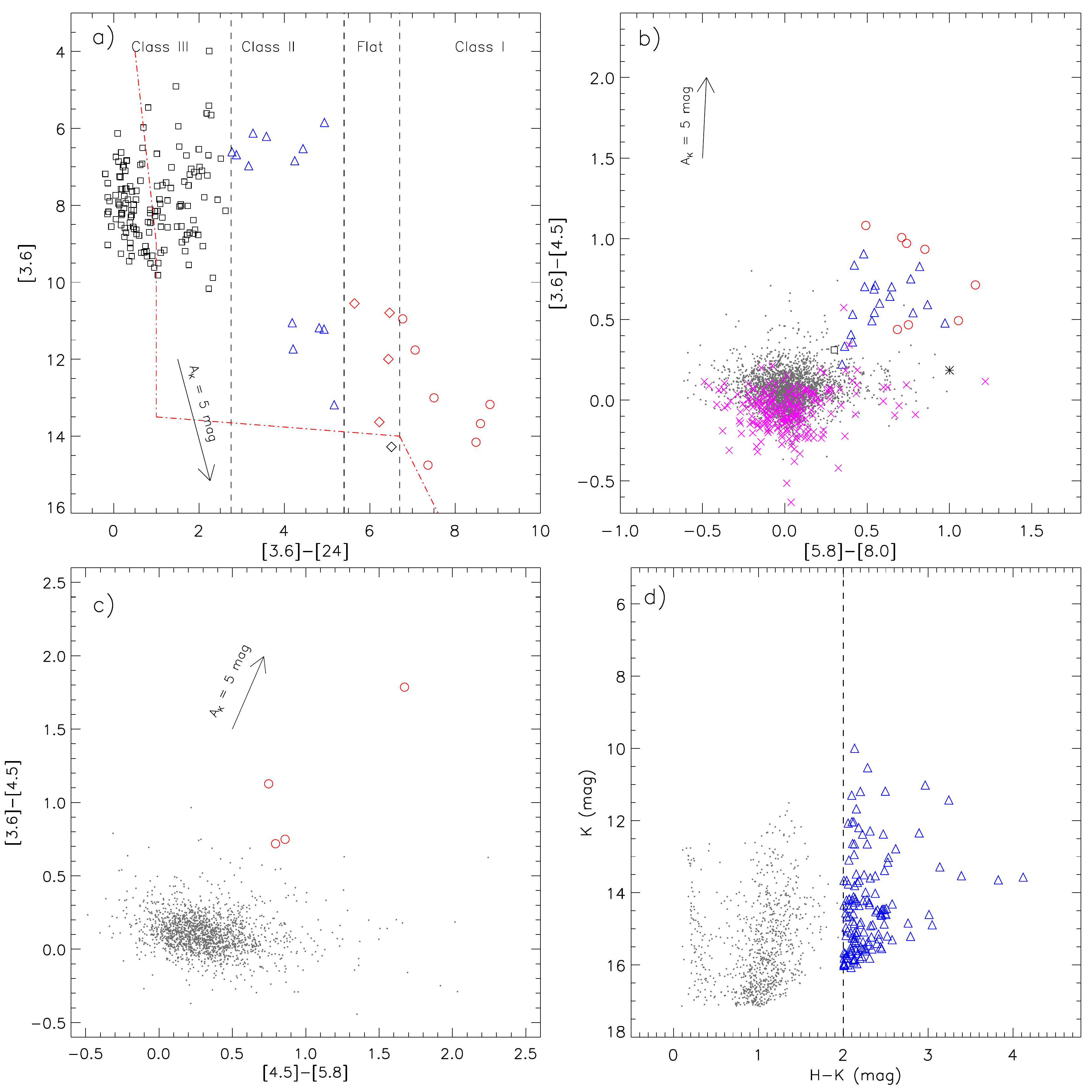}
\caption{\scriptsize Identification of young stellar populations in the region around the G35.6 site.
a) The panel shows a color-magnitude plot ([3.6] $-$ [24] vs [3.6]) of sources detected in the {\it Spitzer} 3.6 and 24 $\mu$m bands (see the text for more details). 
The plot traces YSOs belonging to different evolutionary stages (see dashed lines). 
The zone of YSOs against contaminated candidates (galaxies and disk-less stars) is highlighted by dotted-dashed lines (in red) 
\citep[see][for more details]{rebull11}. Flat-spectrum and Class~III sources are marked by ``$\Diamond$'' and ``$\Box$'' symbols, respectively. 
b) The panel shows a color-color plot ([3.6]$-$[4.5] vs. [5.8]$-$[8.0]) of sources detected in the {\it Spitzer}-GLIMPSE 3.6--8.0 $\mu$m bands. 
The PAH-emitting galaxies and the PAH-emission-contaminated apertures are marked by ``*'' and ``$\times$'' symbols, respectively (see the text). 
A Class~III source is highlighted by ``$\Box$'' symbol in the panel. 
c) The panel shows a color-color plot ([3.6]$-$[4.5] vs. [4.5]$-$[5.8]) of sources detected in the {\it Spitzer}-GLIMPSE 3.6--5.8 $\mu$m bands. 
d) The panel shows a color-magnitude plot (H$-$K/K) of sources selected from the GPS and 2MASS data. 
In all the panels, we show Class~I (red circles) and Class~II (open blue triangles) YSOs. 
In the last three panels, the dots in gray color show the stars with only photospheric emission. 
In the NIR H$-$K/K plot, we have plotted only 901 out of 49399 stars with photospheric emission. 
Due to large numbers of stars with photospheric emission, only some of these stars are randomly marked in the NIR H$-$K/K plot. 
In the first three panels (i.e., color-color diagrams), an extinction vector for A$_{V}$ = 5 mag is drawn 
and is derived using the average extinction laws from \citet{flaherty07}.}
\label{fig6}
\end{figure*}

1. Using the {\it Spitzer} 3.6 and 24 $\mu$m photometric data, \citet{guieu10} used a color-magnitude 
plot ([3.6]$-$[24]/[3.6]) to identify the different stages of YSOs \citep[see also][]{rebull11,dewangan15,dewangan17b}. 
The color-magnitude plot is also utilized to distinguish the boundary of possible contaminants 
(i.e. galaxies and disk-less stars) against YSOs \citep[see Figure~10 in][]{rebull11}. 
We have adopted this scheme in this work and the color-magnitude plot is shown in Figure~\ref{fig6}a.
Following the conditions adopted in \citet{guieu10} and \citet{rebull11}, the boundaries of different stages of 
YSOs and possible contaminants are highlighted in Figure~\ref{fig6}a. 
In Figure~\ref{fig6}a, a total of 164 sources are shown in the color-magnitude space.
We identify 24 YSOs (7 Class~I; 4 Flat-spectrum; 13 Class~II) and 139 Class~III sources. 
Additionally, one Flat-spectrum source is found in the boundary of possible contaminants and is excluded from our selected YSOs. 
In Figure~\ref{fig6}a, the selected Class~I, Flat-spectrum, and Class~II YSOs are highlighted 
by red circles, red diamonds, and blue triangles, respectively.\\  
 
2. Using the {\it Spitzer} 3.6, 4.5, 5.8, and 8.0 $\mu$m photometric data, \citet{gutermuth09} provided 
various schemes to identify the YSOs and also 
various possible contaminants (e.g. broad-line active galactic nuclei (AGNs), PAH-emitting galaxies, shocked emission 
blobs/knots, and PAH-emission-contaminated apertures). Furthermore, the selected YSOs can be further 
classified into different evolutionary stages based on their 
slopes of the SED ($\alpha_{3.6-8.0}$) estimated from 3.6 to 8.0 $\mu$m 
(i.e., Class~I ($\alpha_{3.6-8.0} > -0.3$), Class~II ($-0.3 > \alpha_{3.6-8.0} > -1.6$), 
and Class~III ($-1.6> \alpha_{3.6-8.0} > -2.56$)) \citep[e.g.,][]{lada06}. More details about the YSO classifications 
based on the {\it Spitzer} 3.6--8.0 $\mu$m bands can be found in \citet{dewangan11}. Following the conditions 
given in \citet{gutermuth09} and \citet{lada06}, we have also identified YSOs and various possible contaminants 
in our selected region around the G35.6 site. The {\it Spitzer}-GLIMPSE color-color plot ([3.6]$-$[4.5] vs [5.8]$-$[8.0]) is shown in Figure~\ref{fig6}b. 
Using the {\it Spitzer} 3.6--8.0 $\mu$m photometric data, we find 28 YSOs (8 Class~I; 20 Class~II), 1 Class~III, and 222 contaminants. 
In Figure~\ref{fig6}b, the selected Class~I and Class~II YSOs are highlighted by red circles and blue triangles, respectively.\\  

3. Taking into account the sources having detections in the first three {\it Spitzer}-GLIMPSE bands (except 8.0 $\mu$m band), \citet{hartmann05} and \citet{getman07} use a color-color plot ([4.5]$-$[5.8] vs [3.6]$-$[4.5]) to identify the YSOs. They use color conditions, [4.5]$-$[5.8] $\ge$ 0.7 
and [3.6]$-$[4.5] $\ge$ 0.7, to select protostars.
Using the first three {\it Spitzer}-GLIMPSE bands, we obtain 4 protostars in our selected region (see Figure~\ref{fig6}c). \\ 

4. In this YSO classification scheme, the NIR color-magnitude space (H$-$K/K) is utilized to identify additional YSOs. 
We use a color H$-$K value (i.e. $\sim$2.0) that divides the H$-$K excess sources from the rest of the population.
This color condition is adopted based on the color-magnitude plot of sources from the nearby control field 
(size of the selected region 12$\farcm$1 $\times$ 8$\farcm$7; centered at:  $l$ = 35$\degr$.484; $b$ = $-$0$\degr$.67). 
Using the color H$-$K cut-off condition, we find 142 embedded YSOs in the region probed in this work (see Figure~\ref{fig6}d).\\

All together, the analysis of the photometric data at 1--24 $\mu$m gives a total of 198 YSOs in our selected region around 
the G35.6 site. In Figure~\ref{fig7}a, these selected YSOs are overlaid on the integrated molecular map, 
helping to infer the YSOs belonging to the MCG35.6. We find 59 YSOs in the direction of EMC1, while 15 YSOs are identified toward EMC2. 
In the EMC1, we have found a larger number of YSOs compared to the previously reported work of \citet{paron11} (see Table~2 in their paper). 
Table~\ref{tab4} lists the NIR and {\it Spitzer} photometric magnitudes of the 59 YSOs seen 
toward EMC1, which are identified using the color-magnitude and color-color plots. 
The table also gives the information about the classification stages of YSOs. 
In EMC1, we find noticeable YSOs toward one of the edges of the semi-ring-like morphology. 
Furthermore, several YSOs are also distributed toward the ring-like feature in EMC1.
\subsubsection{Young stellar clusters in MCG35.6}
\label{subsec:surfden}
\begin{figure*}
\epsscale{0.63}
\plotone{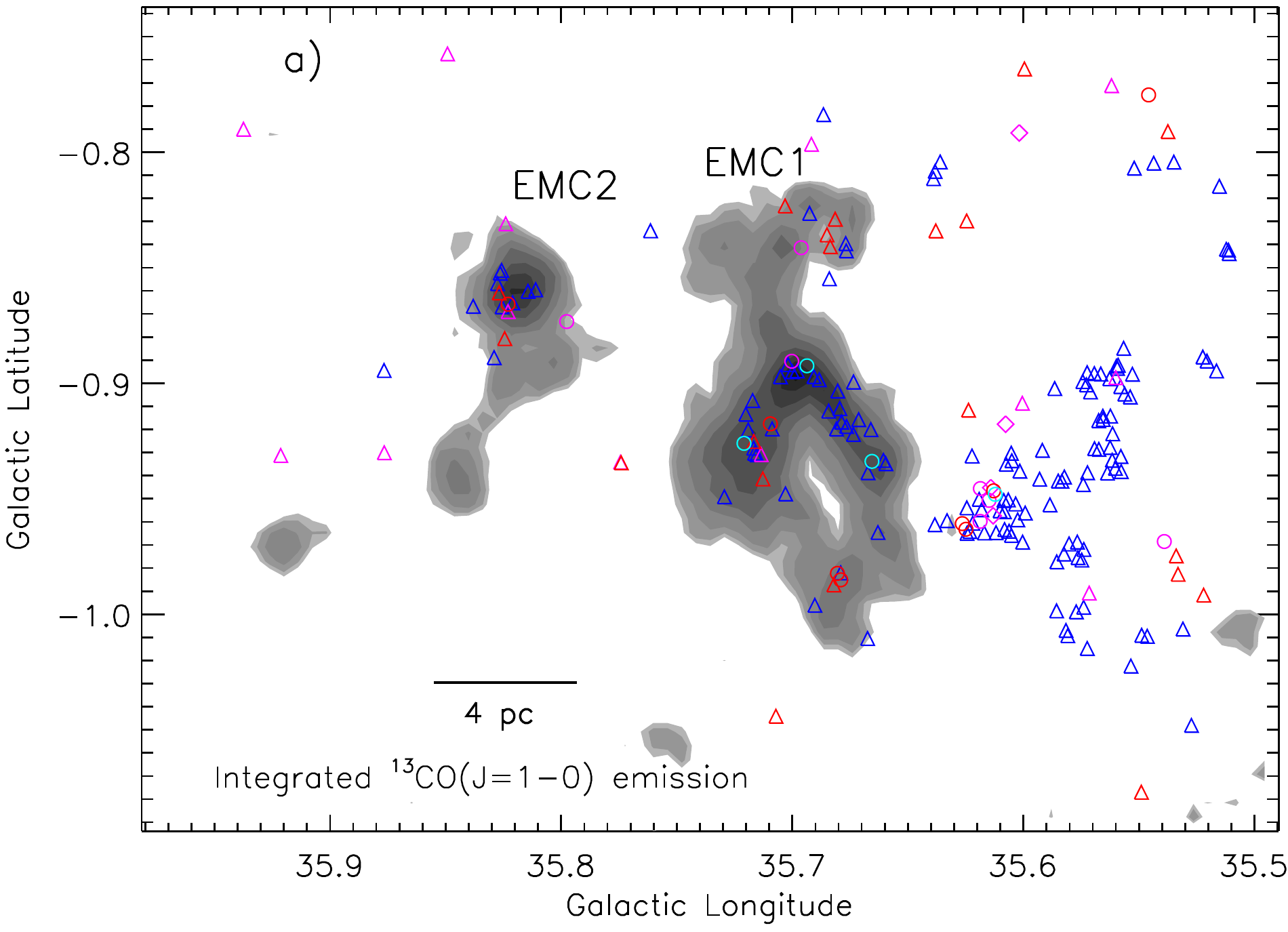}
\epsscale{0.70}
\plotone{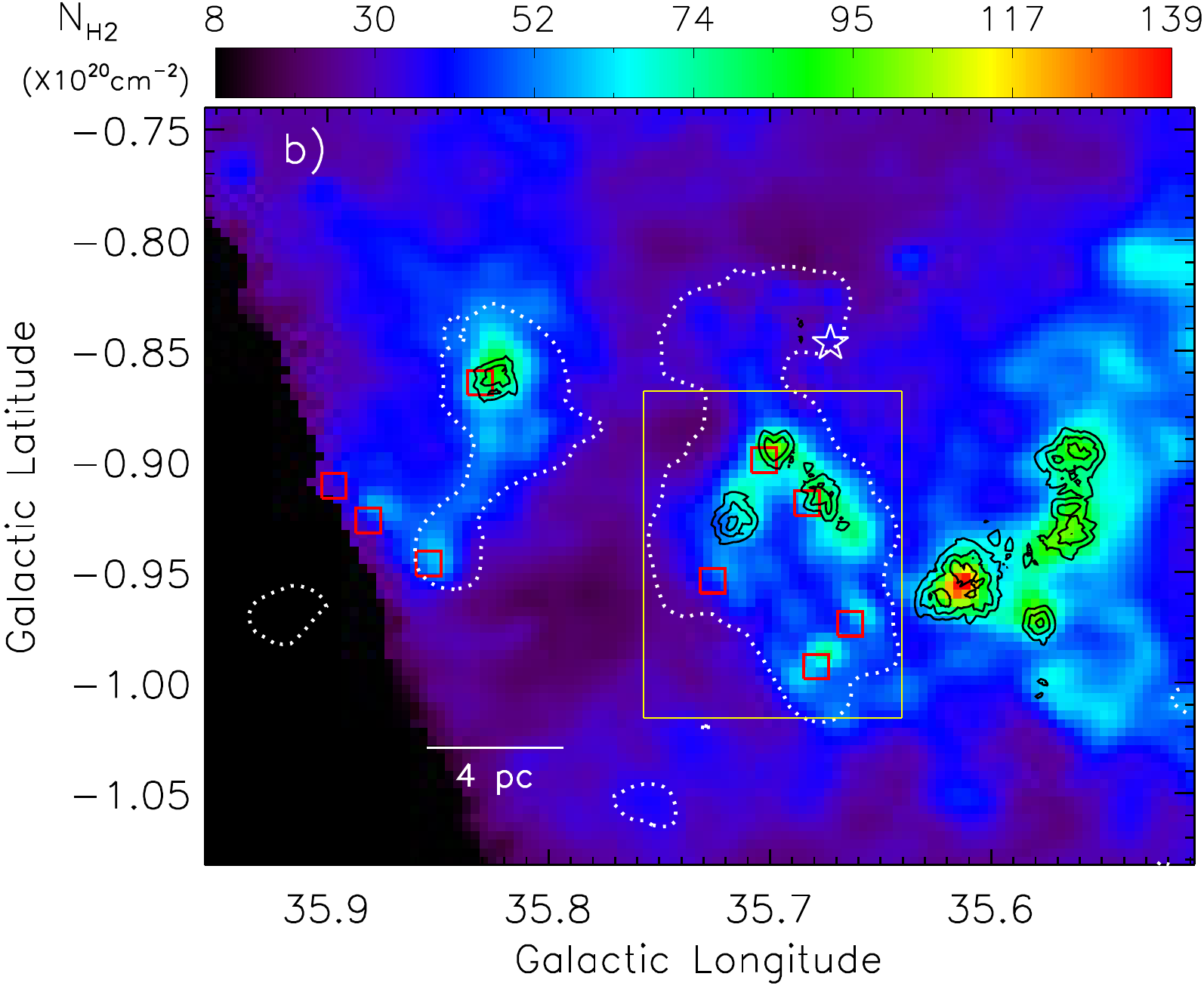}
\caption{\scriptsize Spatial distribution of young stellar populations in the region around the G35.6 site.
a) The selected young stellar populations (Class~I (circles), Flat-spectrum (diamond), and Class~II (triangles)) are shown in the integrated $^{13}$CO map. 
The background map is similar to the one shown in Figure~\ref{fg1}a. 
The YSOs identified using the {\it Spitzer} color-magnitude scheme (i.e. [3.6] $-$ [24] vs [3.6]; see Figure~\ref{fig6}a), 
four {\it Spitzer}-GLIMPSE 3.6-8.0 $\mu$m bands (see Figure~\ref{fig6}b), three {\it Spitzer}-GLIMPSE 
4.5-8.0 $\mu$m bands (see Figure~\ref{fig6}c), and NIR color-magnitude scheme (i.e. H$-$K/K; see Figure~\ref{fig6}d) 
are shown by magenta, red, cyan, and blue color symbols, respectively. 
The molecular map helps us to infer the physical association of YSOs with the MCG35.6. 
b) {\it Herschel} column density map is overlaid with surface density contours (in black) of all the identified YSOs and also {\it Herschel} clumps. 
The surface density contours are drawn at 3, 5, and 10 YSOs/pc$^{2}$, 
increasing from the outer to the inner regions. The background map is similar to the one shown in Figure~\ref{fg4}b. The {\it Herschel} clumps are also highlighted by squares. The solid yellow box encompasses the area shown in Figures~\ref{fig8}a and~\ref{fig8}b}.
\label{fig7}
\end{figure*}

\begin{figure*}
\epsscale{0.5}
\plotone{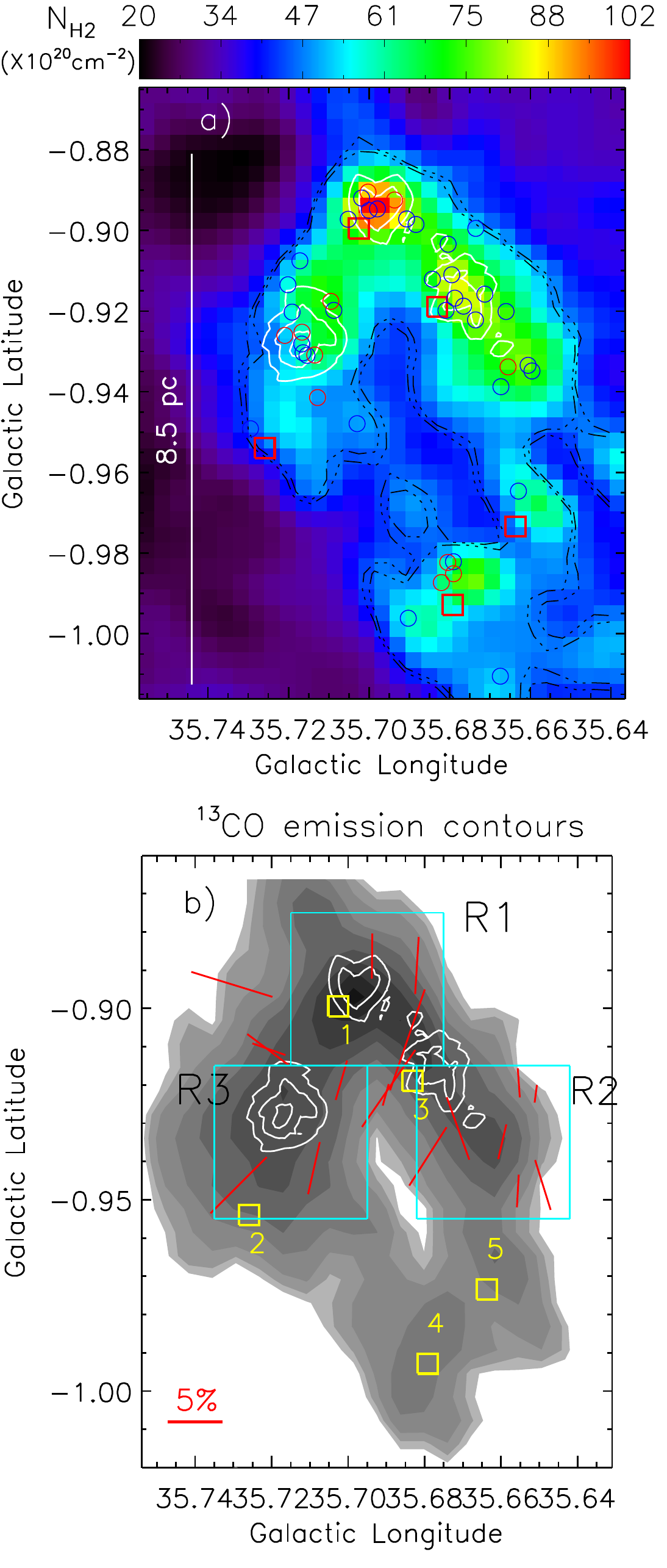}
\caption{\scriptsize Zoomed-in view of the ring-like feature. a) {\it Herschel} column density map is 
superimposed with surface density contours (in white) of all the identified YSOs, {\it Herschel} clumps and selected YSOs. 
b) Overlay of GPIPS H-band polarization vectors (in red) of high quality background stars on the integrated $^{13}$CO 
emission map (see Table~\ref{tab2}). 
The map is also superimposed with surface density contours (in white) of all the identified YSOs and the {\it Herschel} clumps. 
A reference vector of 5\% is highlighted in the lower left corner of the map (see also Figure~\ref{fg5x}b). 
In both the panels, the background maps and other marked symbols are similar to the one shown in Figures~\ref{fg5x}a and~\ref{fg5x}b.}
\label{fig8}
\end{figure*}

In this section, we probe the individual groups or clusters of YSOs based on their spatial distribution and the statistical surface density utility. 
The clusters of YSOs within the MCG35.6 are still unknown. 
In star-forming regions, the surface density map of selected YSOs is often generated using the nearest-neighbour (NN) technique \citep[see][for more details]{gutermuth09,bressert10,dewangan15,dewangan17a,dewangan17b}. 
The surface density map can be obtained by dividing the selected field using a regular grid and extracting 
the surface density of YSOs at each grid point. 
The surface number density at the {\it k$^{th}$} grid point is given 
by $\rho_{k} = (n-1)/A_{k}$ \citep[e.g.][]{casertano85}, where $A_{k}$ denotes 
the surface area defined by the radial distance to the $n$ = 6 NN.
Adopting this procedure, the surface density map of all the selected 198 YSOs has been constructed 
using a 5$\arcsec$ grid and 6 NN at a distance of 3.7 kpc. 
In Figure~\ref{fig7}b, we show the surface density contours of YSOs overlaid on the {\it Herschel} column density map. 
The YSOs density contours are drawn at 3, 5, and 10 YSOs/pc$^{2}$, increasing from the outer to the inner regions. 
The positions of {\it Herschel} clumps are also shown in Figure~\ref{fig7}b. 
The YSO clusters are seen toward {\it Herschel} clumps (i.e., nos. 1, 2, 3, and 8). 
Additionally, noticeable YSOs are also found toward the {\it Herschel} clumps nos. 4 and 5 without any clustering. 
Earlier in Section~\ref{subsec:temp}, the five clumps (i.e., nos. 1, 2, 3, 4, and 5) have been seen at the edges of the ring-like feature (see a solid yellow box in Figure~\ref{fg4}b). 
Figures~\ref{fig8}a and~\ref{fig8}b also show the spatial distribution of YSOs toward the ring-like feature, depicting 
the ongoing star formation activities on the edges of the ring-like feature. 

Star formation activity is also observed toward the clump no. 8 in EMC2. 
Furthermore, we find surface density contours of YSOs away from the $^{13}$CO emission, implying that 
these clusters are not directly associated with MCG35.6. 
\section{Discussion}
\label{sec:disc}
\subsection{H\,{\sc ii} regions in EMC1}
\label{subsec:vmag}
A careful multi-wavelength data analysis reveals two distinct structures (i.e. semi-ring-like and ring-like structures) in the EMC1.
The semi-ring-like structure is associated with the radio source crs1, and is possibly produced by the ionizing feedback of a B0.5V or B0V type star (see Section~\ref{subsec:feed}). 
On the other hand, if the ring-like feature was produced by feedback of a massive star then, 
there should be ionized gas inside the structure produced by a massive star located within the ring-like feature.
Since there are no evidences of the existence of this star or of ionized gas (at the sensitivity of the observed NVSS radio map; 1$\sigma$ $\sim$0.45 mJy beam$^{-1}$), then the ring-like feature is unlikely to be originated by the feedback of a massive star. However, with the sensitivity limit of the NVSS data, one cannot completely rule out the presence of a massive B2-B3 star, or even a cluster of B3 stars, producing the cavity in the ring-like structure. 
Furthermore, the north-western side of the ring-like feature appears to be influenced by the feedback of the massive star linked with the radio source crs2 (see Section~\ref{subsec:feed}).

The photometric analysis of point-like sources reveals a population of YSOs toward the edges of 
the semi-ring-like and ring-like structures 
(see Section~\ref{subsec:phot1}). Hence, star formation activities are traced within EMC1 (see Figure~\ref{fig7}a and also Section~\ref{subsec:phot2}).
The young stellar populations can be originated spontaneously in a given molecular cloud, or the process of collapse and star formation can also be influenced by some external agents such as the expansion of H\,{\sc ii} 
regions \citep[e.g.,][]{elmegreen98}. 
Hence, a triggered star formation scenario (via an expanding H\,{\sc ii} region) appears applicable in the 
G35.6 site. 
The dynamical or expansion ages of the H\,{\sc ii} regions are estimated to be $\sim$0.35--0.5 Myr 
\citep[for n$_{0}$ = 10$^{3}$ cm$^{-3}$; see Section~\ref{subsec:radio}, see also][]{paron11}, which can be treated as the indicative values or the lower limits to the actual ages. 
\citet{evans09} estimated the average lifetimes of Class~I and Class~II YSOs to be $\sim$0.44 Myr and $\sim$1--3 Myr, respectively. Based on the comparison of these typical ages of YSOs and the dynamical ages of the H\,{\sc ii} regions, we cannot completely rule out the formation of Class~I YSOs through the expansion of the H\,{\sc ii} regions in EMC1. 
However, the majority of selected young populations in EMC1 are Class~II YSOs (see Table~\ref{tab4}; 
51 Class~II candidates out of 59 YSOs). Hence, the clusters of YSOs are unlikely to be originated by the expansion of the H\,{\sc ii} regions linked with crs1 and crs2.  
\subsection{Formation and evolution of features in EMC1}
\label{subsec:mag}
The embedded ring-like feature in EMC1 is the most prominent structure evident 
in the $^{13}$CO integrated intensity and {\it Herschel} column density maps (see Figures~\ref{fg5x}a and~\ref{fg5x}b).
The ring-like feature with a face-on view has a central region/cavity devoid of radio continuum emission. 
The structure in the sky of the ring-like feature can be a spherically oblate cloud. 

From a theoretical point of view, the formation of a ring-like feature 
can be explained by the ambipolar diffusion, when the cloud was magnetically subcritical. 
Conditions with low density perturbations 
and thermal pressure allow the cloud to condense along the field lines to 
form the oblate morphologies. When the cloud axis is non-axisymmetric, 
the magnetically mediated cloud condensation forms a ring-like feature 
with dense supercritical cores \citep{fiedler93}. 
\citet{li02} also carried out numerical simulations to examine the evolution of subcritical 
clouds by considering the non-axisymmetric case under thin-disk approximation. 
They reported that a magnetically subcritical cloud fragments into 
multiple magnetically supercritical cores, when the density perturbations are very large. 
Eventually, these supercritical cores lead to the birth of a new generation of stars 
in small clusters through gravitationally dominated ambipolar diffusion.

A detailed analysis of the GPIPS and {\it Planck} polarimetric data 
indicates that the magnetic field is parallel to the major axis of the ring-like feature. 
Our analysis on magnetic field properties further suggests that three 
subregions (i.e. R1, R2, and R3) in the ring-like feature 
are magnetically supercritical and each subregion hosts at least one {\it Herschel} clump (see Section~\ref{subsec:mfrt}).  
We find at least five clumps (i.e., nos. 1, 2, 3, 4, and 5; see Figure~\ref{fg5x}a) at the edges of this ring-like feature which seem to be 
spatially distributed in an almost regularly spaced manner, indicating the fragmentation of the ring-like feature.  
A population of YSOs is found toward all the five clumps including three magnetically supercritical 
clumps (see Section~\ref{subsec:surfden} and Figures~\ref{fig8}a and~\ref{fig8}b). 
One can note that the subregion R2 is seen in the direction of the north-western side of the ring-like feature, 
where the impact of the massive star linked with the radio source crs2 cannot be ignored 
(see Section~\ref{subsec:feed}). However, in the subregion R2, the origin of Class~II YSOs 
through the expansion of the H\,{\sc ii} region is unlikely (see Section~\ref{subsec:vmag}). 

Furthermore, we have also investigated the magnetic field properties in the subregion R4 located at one of the edges of the semi-ring-like morphology (see Section~\ref{subsec:polu}). The field orientation obtained from the GPIPS H-band polarization appears perpendicular to the major axis of the molecular structure linked with the subregion R4 (see Figure~\ref{fg5}a). 
As mentioned earlier, the subregion R4 is magnetically subcritical. 
Additionally, there is no signature of star formation activity in this particular subregion (see Section~\ref{subsec:phot2}). 
This implies that the magnetic field might be dynamically important in the formation and evolution of the semi-ring like feature \citep[e.g.][]{mouschovias78}. 

Our observational results are in agreement 
with the outcomes of the numerical simulations concerning 
the evolution of the magnetically subcritical cloud \citep[see][]{li02}, which fragments into 
multiple magnetically supercritical clumps and further leads to birth of multiple stars/systems or clusters.
\section{Summary and Conclusions}
\label{sec:conc}
In this paper, to probe the physical processes and star formation activity, 
we have carried out an extensive study of the G35.6 site using multi-wavelength data. 
Our analysis has been focused on the distribution of the ionized gas, molecular gas, cold dust, magnetic field, and embedded young populations. The important findings of this work are:\\
$\bullet$ The molecular cloud associated with the G35.6 site (i.e. MCG35.6) is depicted in the velocity range 53--62 km s$^{-1}$. In our selected region around the G35.6 site, 
two extended molecular condensations (i.e., EMC1 and EMC2) are observed.\\ 
$\bullet$ The molecular data trace two molecular components in the direction of the G35.6 site. 
A molecular component associated with EMC1 is depicted in the velocity range 53--59 km s$^{-1}$, 
while the other molecular component associated with EMC2 is observed at 59--62 km s$^{-1}$. 
A careful analysis of the molecular line data shows that these two molecular components are clearly separated in both 
space and velocity, and seem to be unrelated molecular clouds.\\
$\bullet$ In EMC1, the multi-wavelength images reveal the existence of a semi-ring-like feature (associated with the ionized gas emission) and 
an embedded face-on ring-like feature devoid of radio continuum emission in its center. 
The detection/non-detection of the ionized gas emission in the G35.6 site is obtained from the NVSS radio continuum map at 1.4 GHz, which has the sensitivity limit of $\sim$0.45 mJy beam$^{-1}$.\\
$\bullet$ Two compact radio continuum sources (i.e., crs1 and crs2) are associated with EMC1. 
The NVSS 1.4 GHz continuum data analysis reveals that each H\,{\sc ii} region is excited 
by a B0.5V--B0V type star and has a dynamical age of $\sim$0.35--0.5 Myr. \\
$\bullet$ The {\it Herschel} temperature map reveals the heating from the two compact radio continuum sources with temperatures $\sim$22--27~K, 
while the ring-like feature has a temperature range of about 18--20 K. \\   
$\bullet$ In the {\it Herschel} column density map,  at least five massive {\it Herschel} clumps ($M_\mathrm{clump}$ $\sim$740--1420 M$_{\odot}$) 
seem to be spatially distributed in an almost regularly spaced manner along the ring-like feature.\\ 
$\bullet$ The analysis of photometric data at 1--24 $\mu$m reveals a total of 198 YSOs. Most of these YSOs are spatially distributed 
toward the {\it Herschel} clumps in EMC1.\\ 
$\bullet$ Star formation activities have been found toward the clumps seen at the edges of the ring-like feature in EMC1.\\
$\bullet$ The {\it Planck} and GPIPS polarimetric data trace the POS magnetic field parallel to the major axis of the ring-like feature.\\ 
$\bullet$ Using the GPIPS H-band polarimetric data, three magnetically supercritical subregions containing the {\it Herschel} clumps 
are observed in the ring-like feature.\\

Considering our observational findings, the semi-ring-like feature is resultant 
of the ionizing feedback of a massive star linked with the radio source crs1. 
On the other hand, the existence of a ring-like feature cannot be explained 
by the impact of massive star(s). 
Our observational analysis favors the idea that the ring-like feature may have 
formed from a magnetically dominated cloud. 
The results derived using the molecular emission, cold dust emission, and magnetic 
field properties reveal that the ring-like feature contains multiple supercritical 
clumps that can potentially fragment into dense cores, 
eventually leading to the birth of multiple stars/systems or clusters. 
These findings can be linked to the formation and evolution of the 
cloud via the magnetic field mediated mechanism as suggested by Li \& Nakamura (2002).
\acknowledgments
We thank the anonymous reviewer for a critical reading of the manuscript and several useful comments and 
suggestions, which greatly improved the scientific contents of the paper. 
The research work at Physical Research Laboratory is funded by the Department of Space, Government of India. 
This work is based on data obtained as part of the UKIRT Infrared Deep Sky Survey. This publication 
made use of data products from the Two Micron All Sky Survey (a joint project of the University of Massachusetts and 
the Infrared Processing and Analysis Center / California Institute of Technology, funded by NASA and NSF), archival 
data obtained with the {\it Spitzer} Space Telescope (operated by the Jet Propulsion Laboratory, California Institute 
of Technology under a contract with NASA). 
This publication makes use of molecular line data from the Boston University-FCRAO Galactic
Ring Survey (GRS). The GRS is a joint project of Boston University and Five College Radio Astronomy Observatory, 
funded by the National Science Foundation (NSF) under grants AST-9800334, AST-0098562, and AST-0100793.  
This publication makes use of the Galactic Plane Infrared Polarization Survey (GPIPS). 
The GPIPS was conducted using the {\it Mimir} instrument, jointly developed at Boston University and Lowell Observatory
and supported by NASA, NSF, and the W.M. Keck Foundation. 
RD acknowledges CONACyT(M\'{e}xico) for the PhD grant 370405. 
%
%
\begin{table*}
 \tiny
\setlength{\tabcolsep}{0.05in}
\centering
\caption{List of surveys (covering from NIR to radio wavelengths) adopted in this work.}
\label{ftab1}
\begin{tabular}{lcccccr}
\hline 
Survey & Wavelength(s)       & Resolution ($\arcsec$)     & Reference \\
\hline
\hline
Two Micron All Sky Survey (2MASS)                                                 &1.25--2.2 $\mu$m                  & $\sim$2.5          &\citet{skrutskie06}\\
Galactic Plane Infrared Polarization Survey (GPIPS)                                 &1.6 $\mu$m                   & $\sim$1.5          &\citet{clemens12}\\
UKIRT NIR Galactic Plane Survey (GPS)                                                 &1.25--2.2 $\mu$m                   &$\sim$0.8           &\citet{lawrence07}\\ 
{\it Spitzer} Galactic Legacy Infrared Mid-Plane Survey Extraordinaire (GLIMPSE)       &3.6, 4.5, 5.8, 8.0  $\mu$m                   & $\sim$2, $\sim$2, $\sim$2, $\sim$2           &\citet{benjamin03}\\
Wide Field Infrared Survey Explorer (WISE)                                             &3.4, 4.6, 12, 22 $\mu$m                   & $\sim$6, $\sim$6.5, $\sim$6.5, $\sim$12           &\citet{wright10}\\ 
{\it Spitzer} MIPS Inner Galactic Plane Survey (MIPSGAL)                                         &24 $\mu$m                     & $\sim$6         &\citet{carey05}\\ 
{\it Herschel} Infrared Galactic Plane Survey (Hi-GAL)                              &70, 160, 250, 350, 500 $\mu$m                     & $\sim$5.8, $\sim$12, $\sim$18, $\sim$25, $\sim$37         &\citet{molinari10}\\
{\it Planck} polarization data                                                          & 850 $\mu$m & $\sim$294        &\citet{planck14}\\
Galactic Ring Survey (GRS)                                                             & 2.7 mm; $^{13}$CO (J = 1--0) & $\sim$45        &\citet{jackson06}\\
Coordinated Radio and Infrared Survey for High-Mass Star Formation (CORNISH)           &6 cm             & $\sim$1.5        &       \citet{hoare12}    \\     
NRAO VLA Sky Survey (NVSS)                                                             &21 cm            & $\sim$45        &       \citet{condon98}    \\     
\hline          
\end{tabular}
\end{table*}
\begin{table*}
\setlength{\tabcolsep}{0.05in}
\centering
\caption{Physical properties of the {\it Herschel} clumps in our selected star-forming region 
(see Figures~\ref{fg4}b and~\ref{fg4}c). Column~1 gives the IDs assigned to the clumps. Table also lists 
positions, deconvolved effective radius (R$_{c}$), clump mass ($M_\mathrm{clump}$), peak column density (N(H$_{2}$)), 
and extinction ($A_V=1.07 \times 10^{-21}~N(\mathrm H_2)$) \citep{bohlin78}. 
The clumps highlighted with superscript daggers are found at the edges of the ring-like feature (see also Figures~\ref{fg5x} and~\ref{fig8}).}
\label{tab1}
\begin{tabular}{lcccccr}
\hline 
ID & {\it l}       & {\it b}     & R$_{c}$ & $M_\mathrm{clump}$ & peak N(H$_{2}$) & A$_{V}$\\
    &  (degree) & (degree) &  (pc)      &($M_\odot$)   &(10$^{21}$ cm$^{-2}$) &(mag) \\
\hline
\hline
       1$^{\dagger}$  &	 35.702     &	-0.899   &	 1.7	& 1150 &  10       &     11  \\
       2$^{\dagger}$  &	 35.726     &	-0.954   &	 1.5	& 740 &   5.4     &       6  \\
       3$^{\dagger}$  &	 35.683     &	-0.919   &	 1.9	& 1420&   8.5     &       9  \\
       4$^{\dagger}$  &	 35.679     &	-0.993   &	 1.9	& 1275 &   7.7     &       8  \\
       5$^{\dagger}$  &	 35.664     &	-0.973   &	 1.7	&  950 &  6.6    &       7  \\
       6  &	 35.854     &	-0.946   &	 1.5	& 715 &   5.9     &       6  \\
       7  &	 35.881     &	-0.927   &	 1.2	& 450 &   5.6     &       6  \\
       8  &	 35.831     &	-0.864   &	 3.2	&3390 &   9.1     &       10  \\
       9  &	 35.897     &	-0.911   &	 0.8	& 175 &   4.2    &       5  \\
\hline          
\end{tabular}
\end{table*}
\begin{table*}
 \small
\setlength{\tabcolsep}{0.1in}
\centering
\caption{2MASS photometric and GPIPS H-band polarimetric parameters of 17 stars seen toward the ring-like feature. 
Table lists 2MASS magnitudes (J-, H-, and K- bands),  linear polarization (P), uncertainty in P (sP), 
Galactic position angle (GPA), and uncertainty in position angle (sPA) (see Figure~\ref{fg5x}b).}
\label{tab2}
\begin{tabular}{lccccccccccccccr}
\hline 
 {\it l}       & {\it b}     & J-band      &H-band&K-band  &P&sP&GPA&sPA    \\
 (deg)       & (deg)     & (mag)        & (mag)      &(mag)        &(\%)& (\%)&(deg)&(deg)  \\
\hline
\hline
35.682	&	-0.888	&	15.54	&	11.97	&	10.22	&	5.23	&	1.77	&	176.41	&	9.69\\
35.655	&	-0.919	&	12.01	&	 9.61	&	 8.47	&	2.67	&	0.33	&	  3.94	&	3.57\\
35.650	&	-0.922	&	11.71	&	10.50	&	10.02	&	1.62	&	0.28	&	172.62	&	5.07\\
35.684	&	-0.908	&	16.63	&	12.98	&	11.27	&	9.76	&	2.09	&	160.99	&	6.14\\
35.659	&	-0.934	&	16.84	&	12.82	&	10.53	&	3.24	&	1.31	&	166.80	&	11.6\\
35.689	&	-0.920	&	15.61	&	12.86	&	11.59	&	8.56	&	2.46	&	145.21	&	8.24\\
35.671	&	-0.931	&	15.37	&	12.67	&	11.43	&	6.07	&	1.72	&	 19.81	&	8.13\\
35.690	&	-0.922	&	11.36	&	 8.81	&	 7.45	&	2.04	&	0.57	&	164.50	&	8.05\\
35.649	&	-0.946	&	15.81	&	12.03	&	10.31	&	4.77	&	1.06	&	 17.35	&	6.37\\
35.655	&	-0.947	&	12.56	&	10.12	&	 9.00	&	2.98	&	0.23	&	176.77	&	2.23\\
35.679	&	-0.938	&	13.70	&	12.32	&	11.55	&	6.33	&	1.45	&	147.44	&	6.59\\
35.693	&	-0.886	&	15.55	&	11.76	&	 9.79	&	4.12	&	1.58	&	179.38	&	11.01\\
35.730	&	-0.893	&	14.50	&	12.78	&	11.92	&	7.77	&	2.61	&	 72.90	&	 9.63\\
35.701	&	-0.918	&	13.60	&	11.85	&	11.15	&	3.75	&	1.49	&	163.90	&	11.43\\
35.720	&	-0.910	&	14.65	&	11.94	&	10.66	&	3.21	&	1.45	&	 70.58	&	12.93\\
35.708	&	-0.941	&	14.00	&	11.42	&	10.23	&	4.88	&	1.17	&	167.34	&	 6.90\\
35.728	&	-0.946	&	15.20	&	12.57	&	11.35	&	7.27	&	3.28	&	134.94	&	12.91\\
\hline          
\end{tabular}
\end{table*}
\begin{table*}
\setlength{\tabcolsep}{0.05in}
\centering
\caption{Physical properties of four subregions selected in EMC1 (see Figure~\ref{fg5}a). 
Table lists subregions, positions ({\it l} and {\it b}), PA dispersion ($\alpha$), velocity dispersion ($\sigma _{v}$), volume density ($n_{H_{2}}$), 
$<B_{pos}>$, magnetic pressure ($P_{mag}$ = $B_{pos}^2/8\pi$), and normalized mass-to-flux ratio ($\overline{M/\Phi_{B}}$).}
\label{tab3}
\begin{tabular}{lccccccccr}
\hline 
subregion & {\it l}        & {\it b}     & $\alpha$    & $\sigma _{v}$ & $n_{H_{2}}$  &$<B_{pos}>$ & $P_{mag}$& $\overline{M/\Phi_{B}}$\\
            &  (degree) & (degree) &  (degree)                                & (km s$^{-1}$)                               & (cm$^{-3}$)                             &($\mu$G) & ($\times$ 10$^{-11}$ dynes cm$^{-2}$)& \\
\hline
\hline
 1 & 35.695 & -0.895 &  9.9 & 0.79 & 469 & 21.5 & 1.8 & 1.5\\
2 & 35.665 & -0.935 & 18.1 & 0.73 & 465 & 11.0 & 0.48 & 3.0\\
3 & 35.715 & -0.935 & 17.8 & 0.78 & 407 & 11.0 & 0.48 & 2.6\\
4 & 35.725 & -0.845 &  7.0 & 0.68 & 275 & 20.0 & 1.6 & 0.4\\
\hline          
\end{tabular}
\end{table*}
\begin{table*}
\setlength{\tabcolsep}{0.05in}
\centering
\caption{NIR and MIR photometric magnitudes are listed for 59 YSOs (8 Class~I and 51 Class~II) distributed toward EMC1.
These YSOs are identified using the color-magnitude and color-color schemes.
The {\it Spitzer} color-magnitude scheme ([3.6] $-$ [24] vs [3.6]), 
four GLIMPSE bands color-color scheme ([3.6]$-$[4.5] vs. [5.8]$-$[8.0]), 
three GLIMPSE bands color-color scheme ([3.6]$-$[4.5] vs. [4.5]$-$[5.8]), and 
NIR color-magnitude scheme (H$-$K/K) are referred to as sh1, sh2, sh3, and sh4, respectively.}
\label{tab4}
\begin{tabular}{lcccccccccccccr}
\hline 
ID& YSO    & {\it l}   & {\it b}     & J    & H & K  &3.6 $\mu$m & 4.5 $\mu$m& 5.8 $\mu$m& 8.0 $\mu$m& 24 $\mu$m\\
&   (stage,scheme)         &  (degree) & (degree)    & (mag) & (mag)& (mag) & (mag) & (mag)& (mag) & (mag)& (mag) \\
\hline
\hline
1  &    I,sh1  &  35.696  &    -0.841 &      ---  &    --- &	    ---  &     13.18   &    11.30 &    9.67 &	  8.56 &       4.36 \\
2  &    I,sh1  &  35.700  &    -0.891 &      ---  &    --- &	  13.45  &     10.95   &     9.76 &    8.86 &	  8.32 &       4.18 \\
3  &    II,sh1  &  35.713  &    -0.931 &      ---  &    --- &	  13.52  &     11.18   &    10.30 &    9.74 &	  9.23 &       6.37 \\
4  &    I,sh2  &  35.709  &    -0.918 &      ---  &    --- &        ---  &     13.58   &   13.15  &   11.98 &	  11.30&	--- \\
5  &    I,sh2  &  35.680  &    -0.982 &      ---  &  11.46 &  	   9.54  &	7.42   &    6.71  &    5.99 &	   4.83&	--- \\
6  &    I,sh2  &  35.679  &    -0.985 &      ---  &    --- &  	  14.58  &     11.91   &   10.94  &   10.26 &	   9.52&	--- \\
7  &   II,sh2  &  35.681  &    -0.829 &    15.83  &  13.62 &  	  12.49  &     11.28   &   10.69  &   10.05 &	   9.18&	--- \\
8  &   II,sh2  &  35.684  &    -0.841 &      ---  &    --- &	    ---  &     11.55   &   11.05  &   10.49 &	   9.96&	--- \\
9  &   II,sh2  &  35.685  &    -0.836 &      ---  &  15.05 &  	  13.74  &     12.10   &   11.56  &   11.06 &	  10.51&	--- \\
10 &   II,sh2  &  35.703  &    -0.823 &    16.51  &  14.15 &  	  12.97  &     12.08   &   11.60  &   11.26 &	  10.29&	--- \\
11 &   II,sh2  &  35.682  &    -0.987 &      ---  &  14.68 &  	  13.58  &     11.71   &   11.03  &   10.65 &	  10.11&	--- \\
12 &   II,sh2  &  35.717  &    -0.925 &    15.78  &  13.61 &  	  12.11  &     10.11   &    9.57  &    8.99 &	   8.21&	--- \\
13 &   II,sh2  &  35.713  &    -0.941 &    15.71  &  14.11 &  	  13.15  &     11.80   &   11.20  &   10.72 &	  10.14&	--- \\
14 &   II,sh2  &  35.638  &    -0.834 &      ---  &    --- &	    ---  &     13.36   &   12.53  &   12.15 &	  11.33&	--- \\
15 &    I,sh3  &  35.666  &    -0.934 &      ---  &    --- &	    ---  &     14.95   &   13.17  &   11.49 &	  11.34&	--- \\
16 &    I,sh3  &  35.694  &    -0.892 &      ---  &    --- &  	  14.27  &     12.43   &   11.68  &   10.82 &	    ---&	--- \\
17 &    I,sh3  &  35.721  &    -0.926 &      ---  &  14.99 &  	  13.38  &     10.98   &    9.86  &    9.11 &	   8.36&	--- \\
18 & II,sh4  &      35.661   &   -0.933     &  17.91	&   14.13   &	 12.07 &      10.73   &    10.59    &	10.27 &      10.45   &      --- \\
19 & II,sh4  &      35.671   &   -0.916     &  18.62	&   14.78   &	 12.65 &      11.33   &    11.18    &	10.92 &        ---   &      --- \\
20 & II,sh4  &      35.703   &   -0.948     &  20.39	&   17.77   &	 15.56 &      14.38   &    14.13    &	  --- &        ---   &      --- \\
21 & II,sh4  &      35.677   &   -0.919     &	 ---	&   17.93   &	 14.89 &      12.77   &    12.51    &	  --- &        ---   &      --- \\
22 & II,sh4  &      35.639   &   -0.811     &  18.51	&   14.60   &	 12.38 &      10.91   &    10.74    &	10.45 &      10.47   &      --- \\
23 & II,sh4  &      35.638   &   -0.961     &	 ---	&   16.99   &	 14.96 &      14.03   &    13.40    &	  --- &        ---   &      --- \\
24 & II,sh4  &      35.633   &   -0.960     &	 ---	&   15.24   &	 12.34 &      10.82   &    10.52    &	10.08 &      10.12   &      --- \\
25 & II,sh4  &      35.666   &   -0.920     &	 ---	&   16.39   &	 14.34 &      13.04   &    12.91    &	  --- &        ---   &      --- \\
26 & II,sh4  &      35.684   &   -0.912     &	 ---	&   17.74   &	 15.45 &      13.43   &    12.91    &	12.22 &        ---   &      --- \\
27 & II,sh4  &      35.667   &   -0.939     &	 ---	&   16.23   &	 14.11 &      12.69   &    12.62    &	  --- &        ---   &      --- \\
28 & II,sh4  &      35.679   &   -0.917     &	 ---	&   17.96   &	 15.95 &      13.45   &    13.36    &	  --- &        ---   &      --- \\
29 & II,sh4  &      35.679   &   -0.911     &	 ---	&   16.92   &	 13.53 &      11.18   &    10.87    &	10.41 &      11.05   &      --- \\
30 & II,sh4  &      35.681   &   -0.920     &	 ---	&   17.39   &	 15.26 &      13.62   &    13.26    &	  --- &        ---   &      --- \\
31 & II,sh4  &      35.680   &   -0.903     &	 ---	&   14.68   &	 11.43 &       9.34   &     9.07    &	 8.55 &       8.59   &      --- \\
32 & II,sh4  &      35.663   &   -0.965     &	 ---	&   17.71   &	 15.20 &      13.76   &    13.62    &	  --- &        ---   &      --- \\
33 & II,sh4  &      35.674   &   -0.922     &	 ---	&   17.06   &	 14.57 &      12.99   &    12.67    &	11.66 &        ---   &      --- \\
34 & II,sh4  &      35.674   &   -0.900     &	 ---	&   18.05   &	 16.00 &      14.46   &    14.03    &	  --- &        ---   &      --- \\
35 & II,sh4  &      35.720   &   -0.913     &	 ---	&   17.32   &	 15.16 &      13.85   &    13.53    &	  --- &        ---   &      --- \\
36 & II,sh4  &      35.700   &   -0.895     &	 ---	&   18.02   &	 16.02 &	---   &      ---    &	  --- &        ---   &      --- \\
37 & II,sh4  &      35.688   &   -0.899     &	 ---	&   17.99   &	 15.86 &      14.37   &    13.99    &	  --- &        ---   &      --- \\
38 & II,sh4  &      35.636   &   -0.804     &	 ---	&   15.40   &	 12.78 &      11.13   &    10.81    &	10.42 &      10.30   &      --- \\
39 & II,sh4  &      35.690   &   -0.996     &	 ---	&   16.98   &	 14.85 &	---   &      ---    &	  --- &        ---   &      --- \\
40 & II,sh4  &      35.667   &   -1.011     &	 ---	&   17.32   &	 15.08 &      13.69   &    13.53    &	  --- &        ---   &      --- \\
41 & II,sh4  &      35.761   &   -0.834     &  17.80	&   16.89   &	 14.78 &	---   &      ---    &	  --- &        ---   &      --- \\
42 & II,sh4  &      35.677   &   -0.843     &  19.74	&   16.46   &	 14.34 &      12.22   &    11.53    &	10.91 &        ---   &      --- \\
43 & II,sh4  &      35.717   &   -0.908	&  17.33    &	13.40	&    11.30 &	   9.97   &	9.87	&    9.40 &	  9.29   &	--- \\
44 & II,sh4  &      35.705   &   -0.897	&  19.30    &	14.60	&    12.29 &	  10.82   &    10.70	&   10.22 &	 10.19   &	--- \\
45 & II,sh4  &      35.684   &   -0.855	&  18.89    &	14.85	&    12.38 &	  10.91   &    10.58	&   10.25 &	 10.49   &	--- \\
46 & II,sh4  &      35.717   &   -0.928	&  19.24    &	15.67	&    13.66 &	  12.52   &    12.12	&   11.78 &	   ---   &	--- \\
47 & II,sh4  &      35.715   &   -0.931	&  19.48    &	15.06	&    12.94 &	  11.61   &    11.33	&   10.96 &	 10.83   &	--- \\
48 & II,sh4  &      35.719   &   -0.920	&    ---    &	17.89	&    15.56 &	  14.06   &    13.85	&     --- &	   ---   &	--- \\
49 & II,sh4  &      35.677   &   -0.840	&    ---    &	17.85	&    15.68 &	  13.66   &    13.28	&     --- &	   ---   &	--- \\
50 & II,sh4  &      35.702   &   -0.892	&    ---    &	17.68	&    15.54 &	  14.16   &    13.93	&     --- &	   ---   &	--- \\
51 & II,sh4  &      35.698   &   -0.895	&    ---    &	17.97	&    15.85 &	    ---   &	 ---	&     --- &	   ---   &	--- \\
52 & II,sh4  &      35.709   &   -0.920	&    ---    &	17.35	&    14.87 &	  13.48   &    12.85	&     --- &	   ---   &	--- \\
53 & II,sh4  &      35.691   &   -0.897	&    ---    &	17.87	&    15.30 &	  13.65   &    13.12	&     --- &	   ---   &	--- \\
54 & II,sh4  &      35.638   &   -0.808	&    ---    &	17.60	&    14.84 &	  12.94   &    12.42	&   12.11 &	   ---   &	--- \\
55 & II,sh4  &      35.693   &   -0.827	&    ---    &	17.83	&    15.80 &	  13.35   &    13.37	&     --- &	   ---   &	--- \\
56 & II,sh4  &      35.729   &   -0.949	&    ---    &	17.02	&    14.81 &	  13.43   &    13.19	&     --- &	   ---   &	--- \\
57 & II,sh4  &      35.716   &   -0.930	&    ---    &	16.99	&    14.56 &	  13.01   &    12.65	&     --- &	   ---   &	--- \\
58 & II,sh4  &      35.679   &   -0.982	&  15.36    &	14.16	&    12.04 &	  10.25   &	9.92	&    9.25 &	  8.52   &	--- \\
59 & II,sh4  &      35.660   &   -0.935	&  16.84    &	12.82	&    10.54 &	   8.69   &	8.50	&    8.01 &	  7.90   &     6.04 \\
\hline          
\end{tabular}
\end{table*}

\end{document}